\documentclass[prl,twocolumn,superscriptaddress]{revtex4-2}
\pdfoutput=1

\usepackage{graphicx}
\usepackage{subfigure}
\usepackage{dcolumn}
\usepackage{bm}
\usepackage{bbm}
\usepackage{amsthm}
\usepackage{thmtools}
\declaretheorem{theorem}
\usepackage{mathtools}
\usepackage{physics}
\usepackage{binarytree}
\usepackage{tikz,pgfplots}
\usepackage{verbatim}
\usepackage{pgfplots}
\usepackage{placeins}
\usepackage{algorithm2e}
\usetikzlibrary{decorations.pathreplacing}
\usetikzlibrary{arrows.meta}
\usetikzlibrary{graphs}
\usepackage{xcolor}
\usepackage{rotating}
\usepackage{yquant}
\usepackage{tabularray}
\usepackage[colorlinks, linkcolor=red, anchorcolor=blue, citecolor=green]{hyperref}
\RestyleAlgo{ruled}
\pgfplotsset{compat=1.16}
\newtheorem{lemma}{Lemma}
\newtheorem{definition}{Definition}

\begin{document}

\title[Active Quantum Distillation] {Active Quantum Distillation}

\author{Muchun Yang}
\affiliation{Institute of Physics, Beijing National Laboratory for Condensed
  Matter Physics,\\Chinese Academy of Sciences, Beijing 100190, China}
\affiliation{School of Physical Sciences, University of Chinese Academy of
  Sciences, Beijing 100049, China}

\author{D. L. Zhou} \email[]{zhoudl72@iphy.ac.cn}
\affiliation{Institute of Physics, Beijing National Laboratory for Condensed
  Matter Physics,\\Chinese Academy of Sciences, Beijing 100190, China}
\affiliation{School of Physical Sciences, University of Chinese Academy of
  Sciences, Beijing 100049, China}

\date{\today}

\begin{abstract}
  Quantum distillation is a modern technology to decrease the von Neumann entropy of a subsystem by coherent system dynamics. Here we propose an active quantum distillation protocol, in which a bang-bang theme is applied to actively control the coherent dynamics of our system in order to obtain a subsystem with the von Neumann entropy as low as possible. For a bipartite Bosonic system, we derive the analytical expression of the entropy lower bound of one subsystem under any unitary transformation for mixed states with conservation of particles. The lower bound is validated by numerical simulations on the Bose-Hubbard model, where the coherent evolution is controlled by tuning one interaction term of the Hamiltonian.
  Our protocol can be used to decrease the entropy of one subsystem lower than the total bipartite state and increase the number of Bosons or only distill out very few Bosons in the subsystem.
\end{abstract}                         
\maketitle

\textit{Introduction.---}
Decreasing the entropy of a quantum system is an essential task in quantum science and technologies. Low entropy states play  an important role in the study of quantum many-body physics, such as the quantum phase transition~\cite{sachdev_2011,Gadway2022}, topological order states~\cite{zeng2019quantum,PhysRevLett.96.110404} and Bose-Einstein Condensation~\cite{Georgescu2020,RevModPhys.73.307,PhysRevA.97.013605}. To get manifests of those quantum phenomena one must measure some physical quantities on low entropy states. For example, the quantum phase transition is usually accompanied by close of energy gap between their eigenstates, and topological order systems have a topological entanglement entropy on their ground states. Consider a thermal state $\rho = e^{-\beta (H+V)}$ with Hamiltonian $H+V$ and inverse temperature $\beta = 1/T$. If we increase the strength of the potential $V$, it is  equivalent to increase the inverse temperature of the state. Similar idea is also mentioned in quantum virtual cooling~\cite{PhysRevX.9.031013}, which is to square the state to get a half of temperature. Thus the temperature somehow does not capture the essence of the low energy quantumness. However, as a fundamental quantity in thermodynamics and quantum information, low entropy is the most essential manifests of the high purity of a quantum state.

In order to get low entropy states, quantum distillation~\cite{Porto2015,PhysRevA.80.041603,Xia2015} technologies are proposed. Quantum distillation refers to a quantum dynamics that makes the final states smaller samples with purer quantum states than initially present. In the case of fermions, numerical simulation shows that during the expansion and with the strongly repulsive interaction, doublons group together and form a low entropy state~\cite{PhysRevA.80.041603}. In the experiment with a cloud of Bosonic atoms trapped in one-dimensional optical lattices~\cite{Xia2015}, some singlons are observed to quantum distill out of the doublon centre. 

The advancement of quantum simulation techniques based on cold atom experiments has made it possible to experimentally implement quantum distillation protocols. 
Controlling the interaction Hamiltonians for Bose-Hubbard models has been achieved on cold atom experiment platforms~\cite{Salfi2016,Chien2015,Hensgens2017,Schäfer2020,doi:10.1126/science.aal3837,RevModPhys.80.885,doi:10.1126/science.1236362,Hart2015,Winkler2006,PhysRevLett.104.080401,PhysRevLett.110.205301,Schneider2012}. 
And experimentalists can measure the entanglement entropy and mutual information in Bose-Hubbard models by using quantum interference of many-body twins to measure second-order R\'{e}nyi entropy~\cite{Islam2015}.
The Boson-related Hamiltonians can also be simulated on superconducting quantum simulator using a Boson-to-qubit mapping~\cite{peng2023quantum,PhysRevB.93.054116}. 

\begin{figure}[htbp]

    \includegraphics{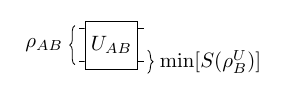}\\
    \centering
    \vspace{-4mm}
    \includegraphics{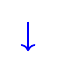}\\
    \centering
    \vspace{-3.5mm}
    \includegraphics[width=7cm,height=4cm]{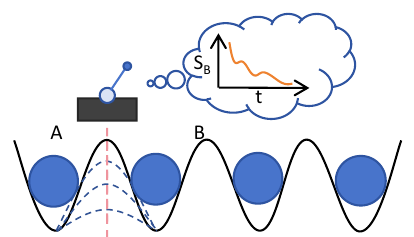}

\caption{Active quantum distillation protocol for decreasing the entropy of system $B$ in the Bose-Hubbard model. The strength of the interaction terms in Hamiltonian is controlled during the time evolution. The total time $T$ is divided into $N$ time steps, each of which is $\delta t$. The interaction strength can be discretely controlled such that the system will evolve along the direction of decreasing $S_B$ to minimum.}
\label{fig1}
\end{figure}

Here we aim at getting low entropy states from a thermal mixed state in Bosonic system through quantum distillation. To be exact, we focus on decreasing the entropy of one subsystem lower than the total system without distilling out many Bosons from this subsystem. Previous quantum distillation protocol based on the expansion from a confined region pure state to other empty lattice. In final state, the region where doublons group together has entropy nearly equal to zero~\cite{PhysRevA.80.041603}. For a initial pure state, the entropy of the final target state will be never less than the total state.
In this Letter we propose the \emph{Active Quantum Distillation}, which is a precise quantum control protocol to decrease the von-Neumann entropy without distilling out many Bosons. For a bipartite Bosonic mixed state $AB$, we derive the analytical expression of the entropy lower bound of subsystem $B$ under a unitary transformation $U_{AB}$. We then construct a unitary evolution by only controlling the interaction Hamiltonian between $B$ and $A$ as illustrated in Fig.~\ref{fig1}. The control protocol is constructed by a simple greedy search algorithm.

Our protocol can be used to group Bosons together or to distill out some Bosons based on different sites configurations and control parameter choices.
In the general case where subsystem $A$ has more than two sites, by choosing appropriate control parameters, we can decrease the entropy of $B$ lower than the entropy of initial total state $AB$, and increase the number of Bosons in $B$ or only distill out very few Bosons. If subsystem $A$ has only one site, we group particles together and decreases the entropy of $\rho_B$ to the entropy of $\rho_{AB}$, which is the lower bound of the entropy of $\rho_B$ under the unitary transformation $U_{AB}$. The essence of the quantum distillation phenomenon is to get a purer state with more Bosons. We explain this quantum distillation phenomenon from the perspective of quantum information theory. Also the protocol stability is tested by numerical simulations.

\textit{Quantum control model.---}
We consider the one-dimensional Bose-Hubbard model with open boundary condition, whose Hamiltonian
\begin{align}\label{H0}
    \hat{H}_0 = -J\sum_{i}^{L-1} (\hat{b}_i^{\dagger} \hat{b}_{i+1} + \text{H.c.}) + \frac{U}{2} \sum_{i=1}^{L} \hat{n}_i(\hat{n}_{i}-1),
\end{align}
where $\hat{b}_i$ ($\hat{b}_i^{\dagger}$) is the annihilation (creation) operator of the Bosonic mode of site $i$, $\hat{n}_i=\hat{b}_i^{\dagger} \hat{b}_{i}$ is the number operator of mode $i$, $J$ is the hopping strength, and $U$ is the onsite interaction energy between two Bosons.
Here we adopt a bang-bang control protocol~\cite{An_2019,PhysRevA.103.012404}, which is widely used in quantum control. In the language of bang-bang control protocol, the total Hamiltonian
\begin{align}
    \hat{H}(t) = \hat{H}_0 + \hat{H}_c(t),
\end{align}
where the Hamiltonian $\hat{H}_0$ specified by Eq.~\eqref{H0} is called the drift Hamiltonian,  and the control Hamiltonian
\begin{align}
    \hat{H}_c(t) = \gamma (t) J(\hat{b}_{l_A}^{\dagger} \hat{b}_{{l_A}+1} + \hat{b}_{{l_A}+1}^{\dagger} \hat{b}_{l_A}),
\end{align}
which plays a role like a switch to adjust the hopping strength between $l_A$th and $(l_A+1)$th sites. Let us divide our system into subsystem $A$ and subsystem $B$, where subsystem $A$ contains the sites from the first to the $l_A$th, and subsystem $B$ contains the remaining sites. Then our central task is to minimize the von Neumann entropy of the subsystem $B$ at a final time $T$ by choosing a suitable $\gamma(t)$, which can be formulated as
\begin{equation}
  \label{eq:1}
   \min_{\gamma(t)} S(\rho_B(T)),
\end{equation}
where \(\rho_B(t)\) is the reduced state of $\rho_{AB}(t)$ ($\rho_{B}(t)= \Tr_A \rho_{AB}(t)$), and the von Neumann entropy  $S(\rho) = -\Tr \rho \ln \rho$. Note that the minimal entropy depends on the initial state $\rho_{AB}(0)$. In our protocol, we take the initial state of the whole system $\rho_{AB}(0)=e^{-\beta H_0}/Z$ with the partition function $Z=\Tr(e^{-\beta H_0})$, where $\beta=\frac{1}{k_BT}$, $k_{B}$ is the Boltzmann constant, and $T$ is the temperature. In most cases, the minimum entropy obtained is smaller than $S(\rho_{AB}(0))$.
We will see in the following sections that, in the task of minimizing the subsystem $B$ entropy, our protocol is nearly equivalent to that capable of applying any global unitary transformations when $l_A = 1$. And for general cases when $l_A > 1$, our protocol has the appropriate capacities to decrease the entropy of $B$ to be less than $S(\rho_{AB}(0))$ without losing many Bosons.

\textit{Analytical lower bound of subsystem entropy.---}
In this section we derive the analytical expression of the lower bound of one subsystem entropy under any unitary transformation. This lower bound is universal for all Bosonic system but not restricted in the Bose-Hubbard model, since in the derivation we only restrict the total particle number to be conserved and all modes are the same after the unitary transformation. In other words, let $\hat{N}_{AB} = \sum_{i=1}^{L} \hat{n}_i$, in this section we derive the lower bound under any unitary transformation which is generated by a Hamiltonian commuted with $\hat{N}_{AB}$. 

Suppose that $\rho_{AB}=\rho_{AB}(0)$ is a bipartite state with total particle number $N_{AB}$. The number of sites in $A$ is $l_A$, and the number of sites in $B$ is $l_B$ with $l_A + l_B = L$.
We take the occupation number bases $\{ |n_1, n_2, \dots ,n_L \rangle \}$, which are defined as $\hat{n_i} |n_1, n_2, \dots ,n_L \rangle = n_i |n_1, n_2, \dots ,n_L \rangle $, and satisfies $\sum_{i=1}^L n_i = N_{AB}$. The reduced density operator of $A$ and $B$ is $\rho_{A} = \Tr_B \rho_{AB}$ and $\rho_{B} = \Tr_A \rho_{AB}$. The occupation number bases for $\rho_{A}$ are $\{ |n_1, n_2, \dots ,n_{l_A }\rangle_A \}$, and for $\rho_{B}$ are $\{ |n_{{l_A}+1}, \dots ,n_L \rangle_B \}$. The number operator of $A$ is $\hat{n}_A = \sum_{i=1}^{l_A} \hat{n}_i$, and the number operator of $B$ is $\hat{n}_B = \sum_{i={l_A}+1}^{L} \hat{n}_i$. Let $\hat{n}_A |n_1, n_2, \dots ,n_{l_A} \rangle_A = n_A |n_1, n_2, \dots ,n_{l_A} \rangle_A $, and $\hat{n}_B |n_{{l_A}+1},\dots ,n_L \rangle_B = n_B |n_{{l_A}+1},\dots ,n_L \rangle_B $. 

Both $\rho_{A}$ and $\rho_B$ are block-diagonal matrices, where different block has bases corresponding to different $n_A$ and $n_B$. So we can write $\rho_{A} = \oplus_{i=0}^{N_{AB}} \rho_{A,i}$ and $\rho_{B} = \oplus_{i=0}^{N_{AB}} \rho_{B,i}$. The dimension of each $\rho_{A,i}$ is $d_{A,i} = (i+{l_A}-1)!\big/\big[i!({l_A}-1)!\big]$, where $i=N_{AB},\dots,1,0$. And the dimension of each block in $\rho_B$ is $d_{B,i} = (i+{l_B}-1)!\big/\big[i!({l_B}-1)!\big]$, where $i=0,1,\dots,N_{AB}$. $d_{A,i}$ and $d_{B,i}$ satisfy $\sum_{i=0}^{N_{AB}} d_{A,i} \times d_{B,N_{AB}-i} = d_{AB}$, where $d_{AB}$ is the dimension of $\rho_{AB}$.

Note that, for disentangled subsystems $A$ and $B$, total system $AB$ is a direct sum of the sector labeled by $n_A$ and $n_B$ (also known as superselection rule~\cite{PhysRevA.69.052326}), 
\begin{align}\label{rho_AB_blocks}
    \rho_{A} \otimes \rho_{B} = \mathop{\oplus}\limits_{n_A+n_B=N_{AB}} \rho_{n_A,n_B},
\end{align}
where $\rho_{n_A,n_B}$ is a sector whose subsystem $A$ only has bases with $n_A$ and subsystem $B$ only has bases with $n_B$. There exists unitary transformations that can entangle $A$ and $B$,
\begin{align}\label{rho_AB_blocks_and_non_diagonal}
    U\rho_{A} \otimes \rho_{B} U^{\dagger}=& \mathop{\oplus}\limits_{n_A+n_B=N_{AB}} \Tilde{\rho}_{n_A,n_B}\nonumber \\
                                            &+ \text{\{non-block-diagonal elements\}},
\end{align}
 i.e., non-block-diagonal elements means the entanglement between two subsystems. Denote the Hilbert space of $\rho_{n_A,n_B}$ as $\mathcal{H}_{{n_A,n_B}}$, and denote system $A$ and $B$ in $\mathcal{H}_{{n_A,n_B}}$ as $\mathcal{H}_{A,n_A}$ and $\mathcal{H}_{B,n_B}$.
 In each $\rho_{n_A,n_B}$ the Hilbert space has tensor product structure, i.e. $\mathcal{H}_{n_A,n_B} = \mathcal{H}_{A,n_A} \otimes \mathcal{H}_{B,n_B}$.

Before going into our Theorem, we omit the non-block-diagonal elements and rearrange the order of the sectors in Eq.~\eqref{rho_AB_blocks}. The rearranged sectors are 
\begin{align} \label{rho_k}
   \rho_{AB} = \oplus_{k=0}^{N_{AB}} \rho_{k}, 
\end{align}
such that the dimension of $A$ of each $\rho_k$ is in decreasing order, i.e.,  $d_{A,k} \geq d_{A,k+1}$. For each $\rho_k$, the dimension of $AB$ is $d_k$.

\begin{theorem}
\label{thm1}
Suppose the eigen-decomposition of the bipartite particle number conservation state $\rho_{AB}^0$ is $\rho_{AB}^0 = \sum_{j=1}^{d_{AB}} p_j |\psi_j\rangle \langle \psi_j|$. We assume these eigenvalues are in decreasing order, i.e., $p_j \geq p_{j+1}$. Divide all eigenvalues ${p_j}$ into $N_{AB}$ groups $\varrho_k$, $k = 1,2,\dots,N_{AB}$. The number of eigenvalues in group $\varrho_k$ is $d_k$. Denote the $l$th eigenvalue in the $k$th sector as $p_{k,l}$. The division satisfies, 
\begin{align}
    &\text{(i) inside the}~k\text{-}th~\text{group}, p_{k,l} \geq p_{k,l+1}, \nonumber \\ 
    &\text{(ii) for all}~l,l'~\text{in}~k\text{-}th~\text{and}~(k+1)\text{-}th~\text{groups}, p_{k,l} \geq p_{k+1,l'}. \nonumber
\end{align}

Let 
\begin{align}\label{qkb}
    q_{k,b} = \sum_{a=1}^{d_{A,k}} p_{k,(b-1)d_{A,k} + a }~~~.
\end{align}

By performing a unitary operation $U$ on the AB, the minimal entropy of the subsystem B is 

\begin{align}
    \min_{U} S(\rho_{B}(U)) = -\sum_{k}\sum_{b} q_{k,b} \log_2 q_{k,b}.
\end{align}

In particular, if A only has one site, then the dimension of $\rho_B$ equals to the dimension of $\rho_{AB}^0$, $d_B = d_{AB}$. In this case the optimal $U$ satisfies $\rho_{AB}(U) = \rho_B(U)$, and the condition (i) and (ii) can be ignored. Thus $\min_{U} S(\rho_{B}(U)) = S(\rho^0_{AB})$.
\end{theorem}

\renewcommand{\proofname}{Proof sketch}
\begin{proof}--

    The Lemma we mainly use is for the majorization and von-Neumann entropy~\cite{nielsen2002introduction}: Suppose $\rho$ and $\sigma$ are density operators such that $\rho \prec \sigma$, then $S(\rho) \geq S(\sigma)$. For two quantum state density operator $\rho$ and $\sigma$, $\rho\prec \sigma$ if and only if $\lambda_{\rho} \prec \lambda_{\sigma}$, where $\lambda_{\rho(\sigma)}$ is the list of eigenvalues for operator $\rho$($\sigma$). The majorization between two real vectors $\vec{a} \prec \vec{b}$ with dimension $D$ is defined as (i)$\sum_{i=1}^d a_i \leq \sum_{i=1}^d b_i $, $1\leq d \leq D$, and (ii) $\sum_{i=1}^D a_i = \sum_{i=1}^D b_i = 1$. 
    
    For the case that A only has one site, $l_A = 1$, and $d_B = d_{AB}$. The trace operation actually drops out those non-block-diagonal elements in $\rho_{AB}$, i.e. $\rho_{B} = \rho_{AB} - \text{\{non-block-diagonal elements\}}$. When $U$ satisfies $\rho_{AB} = \rho_B$, non-block-diagonal elements are zeros. If $\rho_{B} \neq \rho_{AB}$, $\rho_{AB}$ has non-zero non-block-diagonal elements. There exists a block-diagonal unitary matrix $U_D$ that can diagonalize $\rho_B$, $U_D\rho_B U_D^{\dagger} = diag\{\lambda_B\}$, where $diag\{\lambda_B\}$ represents a diagonal matrix with diagonal elements being the eigenvalues of $B$. Note that $U_D\rho_{AB} U_D^{\dagger} = diag\{\lambda_B\} + \text{\{non-block-diagonal elements\}}$. By Schur-Horn theorem, $\lambda_{AB} \succ \lambda_B$, therefore $S(B)\geq S(AB)$.

    For general cases that the sites number of subsystem $A$ is larger than one, $l_A >1$, and $d_B < d_{AB}$. We aim to show $\rho_B \prec \rho_B^*$, where $\rho^*_B$ is the optimal state of $B$ in the Theorem~\ref{thm1}, and $\rho_B$ is any other state different from $\rho^*_B$. From the above situation we learn that the non-block-diagonal elements will increase the entropy of one subsystem, so in the following we will only consider the density operator being block-diagonal, and only focus on any one of the sectors. We first note that if the sectors in $\rho_{AB}$ is not diagonal, there exist a doubly stochastic matrix $D$, such that $\lambda_{\rho_B} = D \times \lambda_{ \rho_B^* }$. 
    And this will make $\lambda_{\rho_B} \prec \lambda_{ \rho_B^* }$ and leads to a greater entropy of $B$. We next note that inside one diagonal sector, each eigenvalue of $\rho_B$ in sector $\rho_k$ is obtained from sum of $d_{A,k}$ eigenvalues in $\rho_{k}$. So we want to arrange the largest $d_{A,k}$ eigenvalues of $\rho_{AB}$ into the sector with largest $d_{A,k}$, and so on, until the smallest $d_{A,k}$ eigenvalues of $\rho_{AB}$ for the sector with smallest $d_{A,k}$. Any other order of the $\lambda_{\rho_{AB}}$ in $\rho_{AB}$ leads to $\lambda_{\rho_B} \prec \lambda_{ \rho_B^* }$. 
    
    The full proof in details is given in Supplemental Material~\cite{SM}.
\end{proof}

\begin{table*}[htbp]
  \centering
  \begin{tblr}{|c|ccccccc|}
    \hline
    $(N_{AB}, L)$ &     $(1,3)$& $(2,3)$ &  $(3,3)$ &  $(2,4)$ &  $(3,4)$ &  $(4,4)$  &\\
    \hline
    difference                          &    0.016    &0.055    &     0.039&     0.022&   0.034  & 0.069     &\\
    \hline
    T                     &      20     &  100      &       30&30         &      50  &       40&\\
    \hline
  \end{tblr}
  \caption{The difference between final entropy and theoretical lower bound, and total evolution time $T$ for different size of Bose-Hubbard models with $l_A = 1$. From the table we can see that the total time and errors are parameter-dependent. The total time $T$ is relatively scaling with $O(1)$ but not scaling exponentially.}
\label{tab:bose}
\end{table*}

\begin{figure*}[htbp]

\subfigure[]{
\includegraphics[width=3.5cm,height=3.5cm]{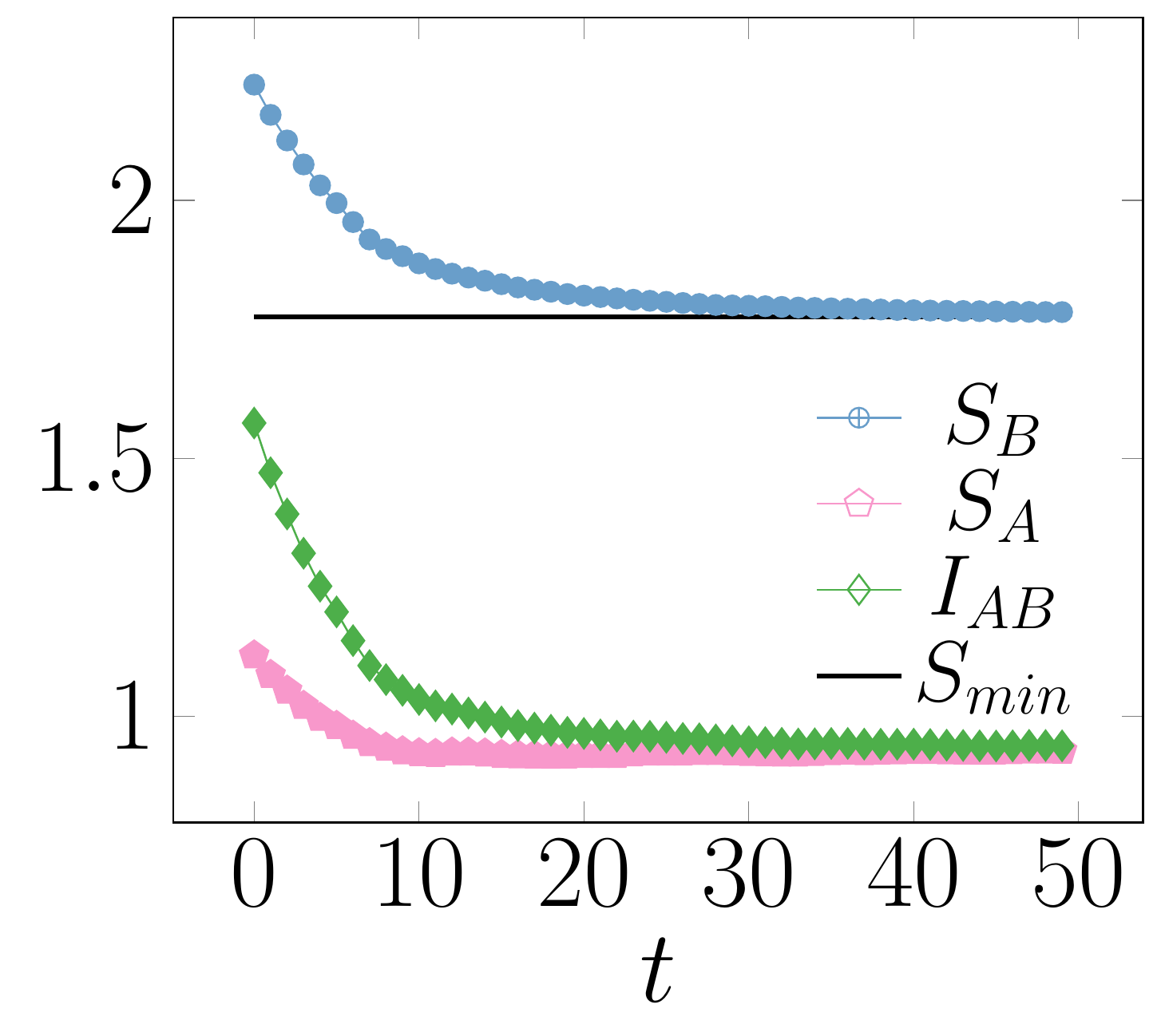}
}
\hspace{-4mm}
\subfigure[]{
\includegraphics[width=3.5cm,height=3.5cm]{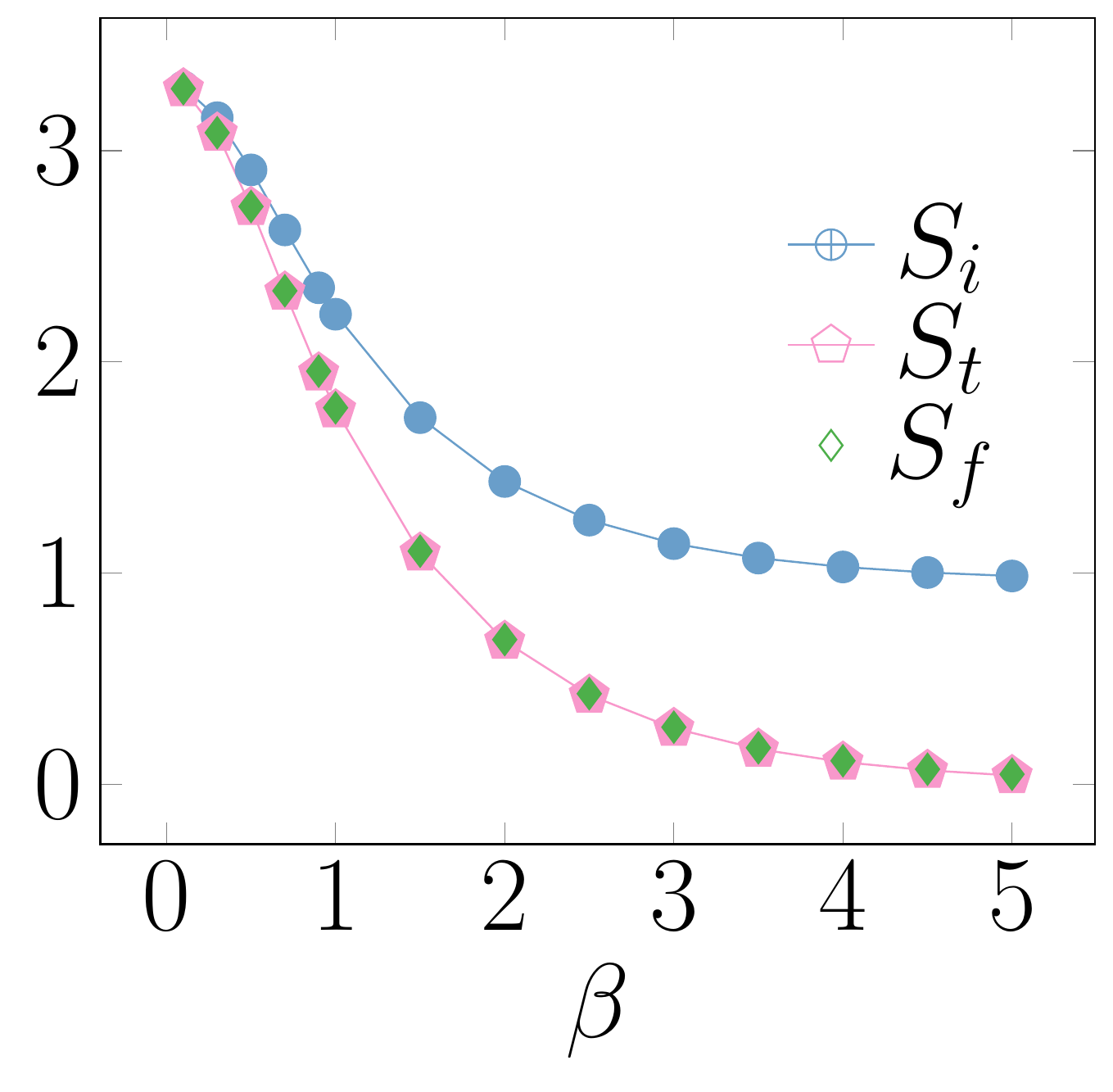}
}
\hspace{-4mm}
\subfigure[]{
\includegraphics[width=3.5cm,height=3.5cm]{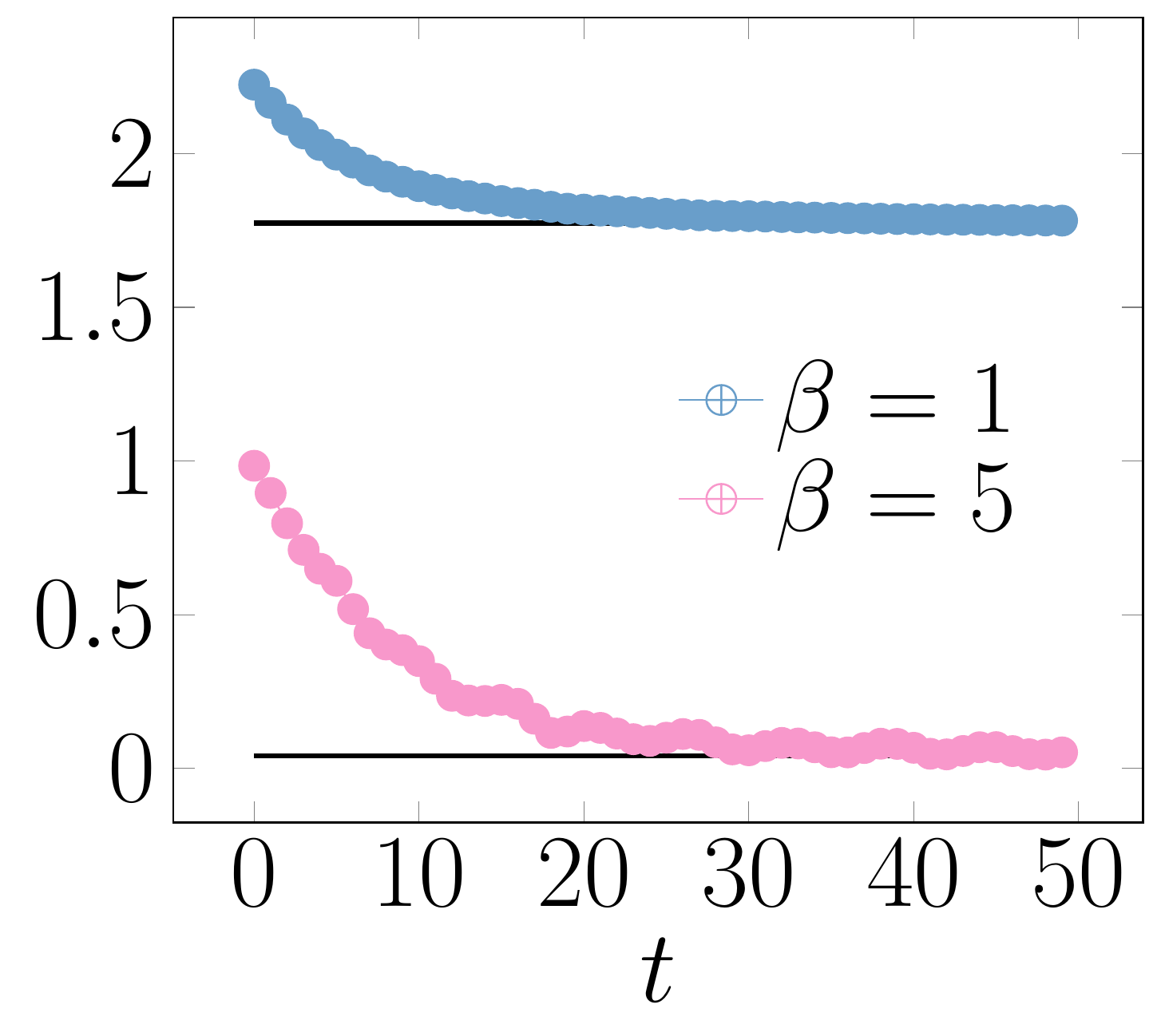}
}
\hspace{-4mm}
\subfigure[]{
\includegraphics[width=3.5cm,height=3.5cm]{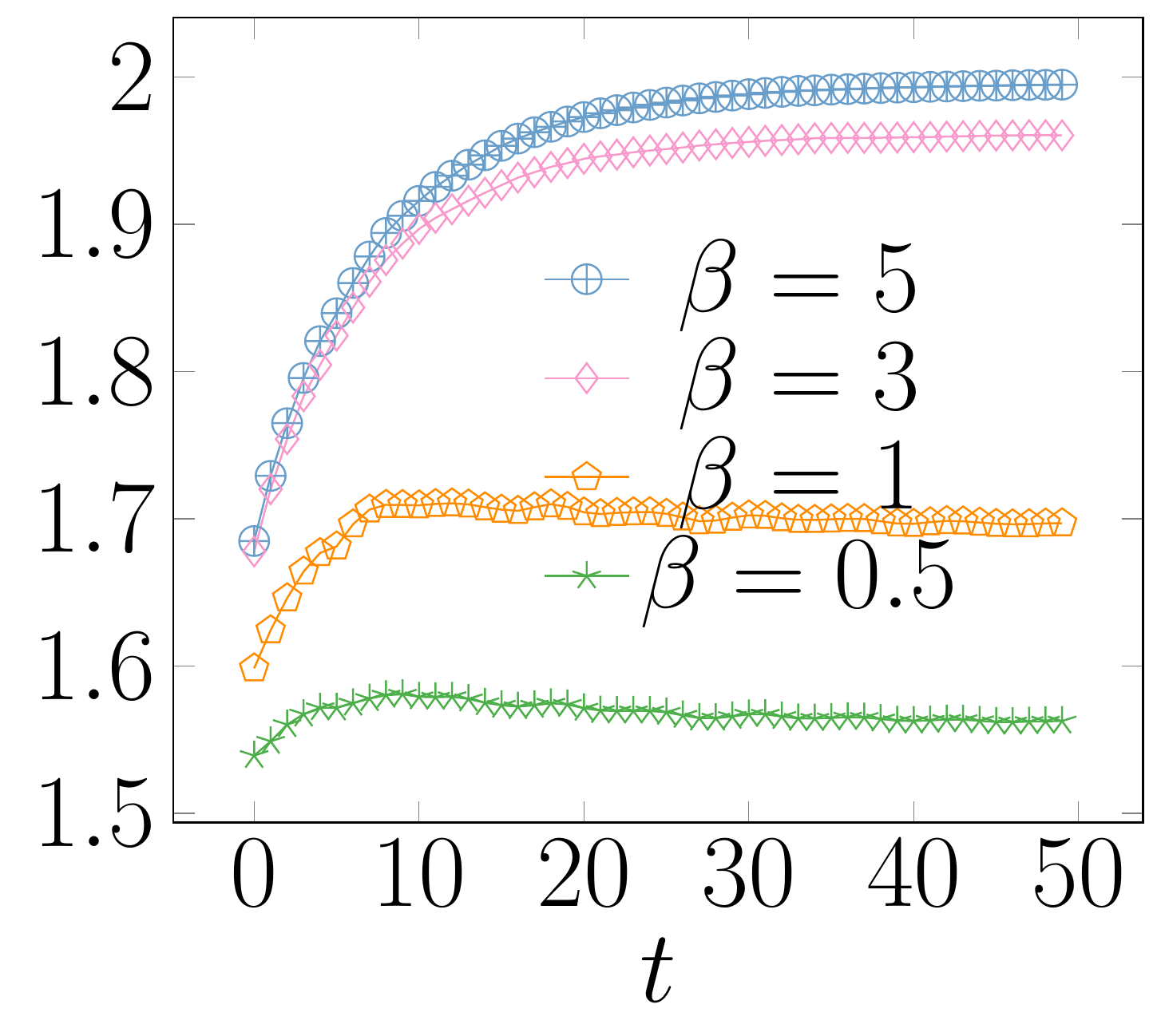}
}
\hspace{-4mm}
\subfigure[]{
\includegraphics[width=3.5cm,height=3.5cm]{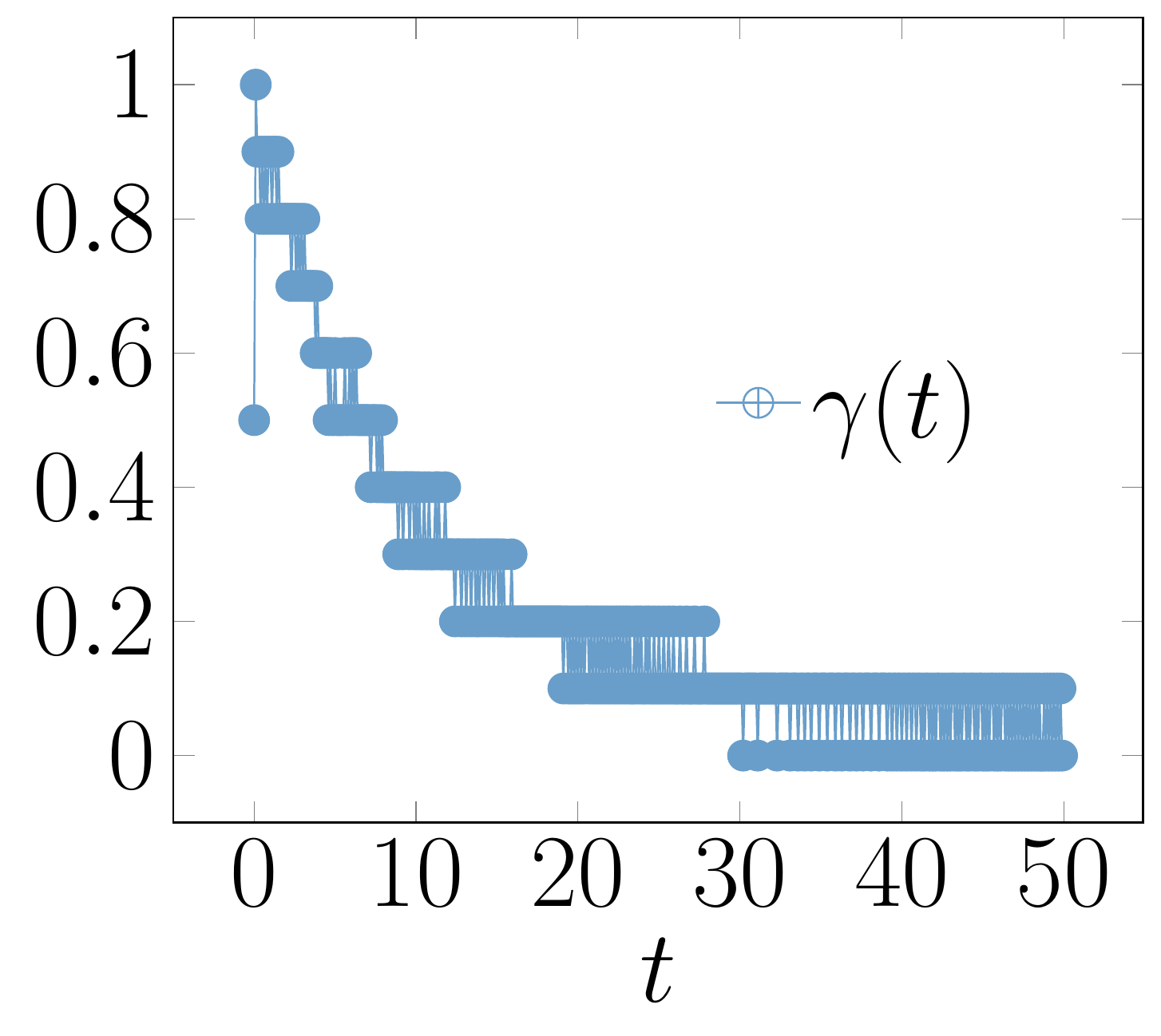}
}
\caption{
The numerical results on the Bose-Hubbard model, $L = 4$, $l_{A} = 1$, $N=2$, $J = U = 1$.
(a) The entropy of subsystem $B$ as the function of time $t$. $\beta =1$, $\delta t = 0.1$.
(b) The final entropy of subsystem $B$ as a function of $\beta$, with other parameters fixed. $S_i$ is the entropy of initial state $\rho_{B}$, and $S_t$ is the entropy theoretical lower bound of $\rho_{B}$.
(c) The entropy of subsystem $B$ as the function of time $t$ for $\beta =1$ and $\beta =5$, the control parameters are identified on model with $\beta =1$ and are tested on model with $\beta =5$.
(d) The number of Bosons $\langle \hat{n}_B \rangle$ in subsystem $B$ as the function of time $t$ for different $\beta$.
(e) The choice of $\gamma$ as the function of $t$. The path of the $\gamma(t)$ tends to be an adiabatic evolution.
}
\label{entropy2}
\end{figure*}

\textit{Numerical Results.---}
The total evolution time $T$ is divided into $N$ identical time steps $\delta t = T/N$. In the $j$th time step, $(j-1) \times \delta t<t< j \times \delta t$, the control parameter $\gamma_{j}^{(k)}$ $(k = 1,\dots,d)$ is selected from $\mathcal{C}(\gamma)$, which is a set of $d$ possible choices. Thus $H_j^{(k)} = H_0 + H_c (\gamma^{(k)}_j)$ and $U_j^{(k)} = e^{-iH_j^{(k)} \delta t}$. The selected $N$ control parameters construct a path $\Gamma = \{\gamma^{(k_1)}_1, \gamma^{(k_2)}_2, \dots ,\gamma^{(k_N)}_N \}$. The total unitary transformation is $U = \prod_{j=0}^{N} U_j$.

We use a greedy algorithm to find the control parameters path. For an initial state $\rho_0$, at the beginning of time step $t_j$, the state is $\rho_{j-1} = \prod_{i=0}^{j-1} U_{i} \times \rho_0$.
 We choose the control parameter in $t_j$ to be 
\begin{align}\label{greedy_algorithm}
    \gamma (t_j) = \underset{\gamma_{j}^{(k)}} {\operatorname{argmin}}~S(\rho_{B,j}^{{(k)}}),
\end{align}
where $\rho_{j}^{(k)} = U_j^{(k)} \times \rho_{j-1}$, and $\rho_{B,j}^{{(k)}} = \tr_A \rho_{j}^{(k)}$. If  $S(\rho_{B,j}^{(1)}) = S(\rho_{B,j}^{(2)}) = \dots = S(\rho_{B,j}^{(d)})$, then we choose
\begin{align}\label{greedy_algorithm_1}
    \gamma (t_j) = \underset{\gamma_{j}^k} {\operatorname{argmax}}~\gamma_{j}^k.
\end{align}
In the following numerical simulations, we choose $\mathcal{C}(\gamma) = \{1,~0.9,~0.8,~\dots,~0.1,~0\}$. In actual numerical simulation we first choose the value of $\delta t$, and the value of $T$ is chosen to make $S(\rho_B)$ converge to its minimum.

\begin{figure*}[htbp]

\subfigure[]{
\includegraphics[width=4.4cm,height=4.4cm]{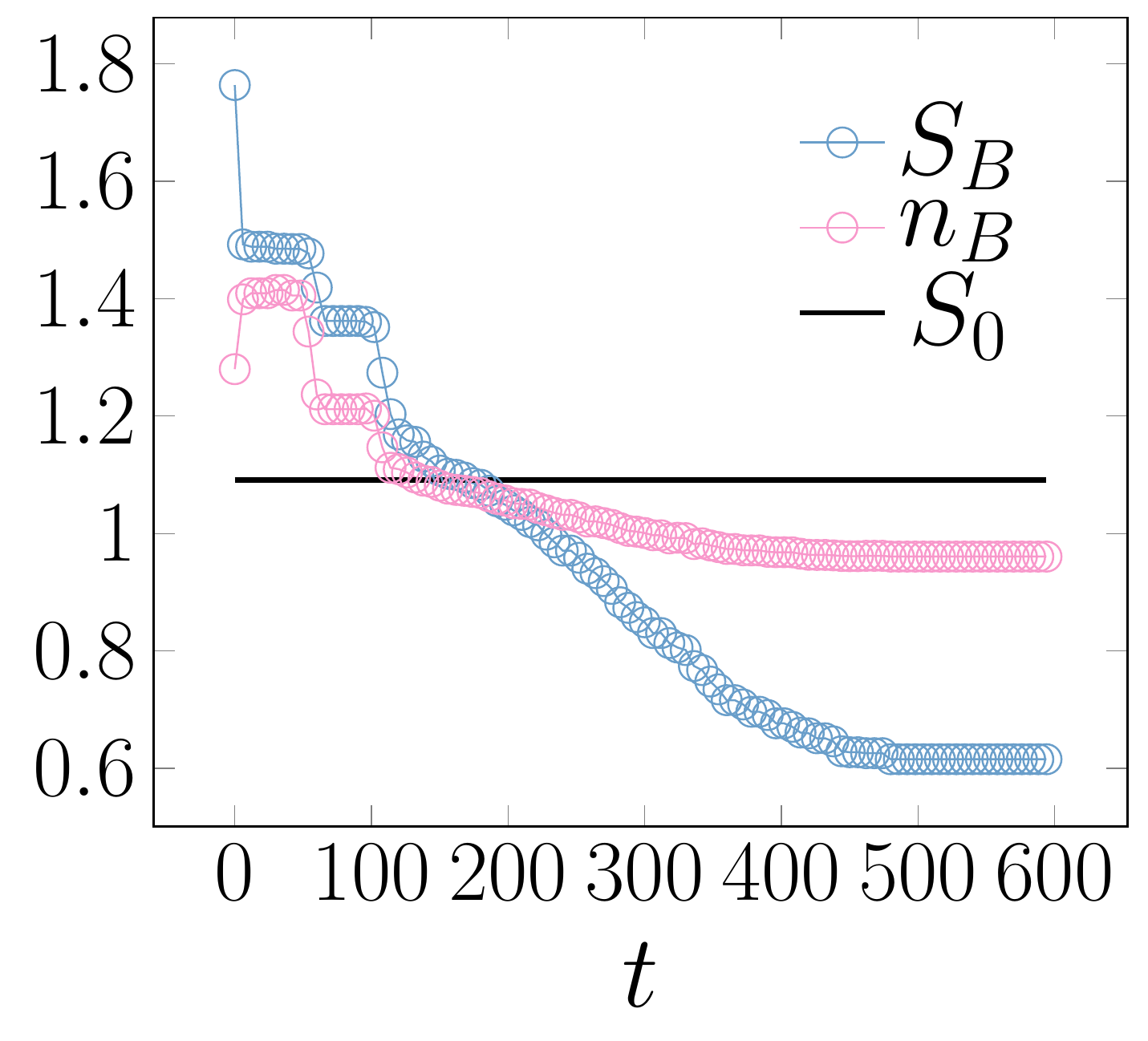}
}
\hspace{-4mm}
\subfigure[]{
\includegraphics[width=4.4cm,height=4.4cm]{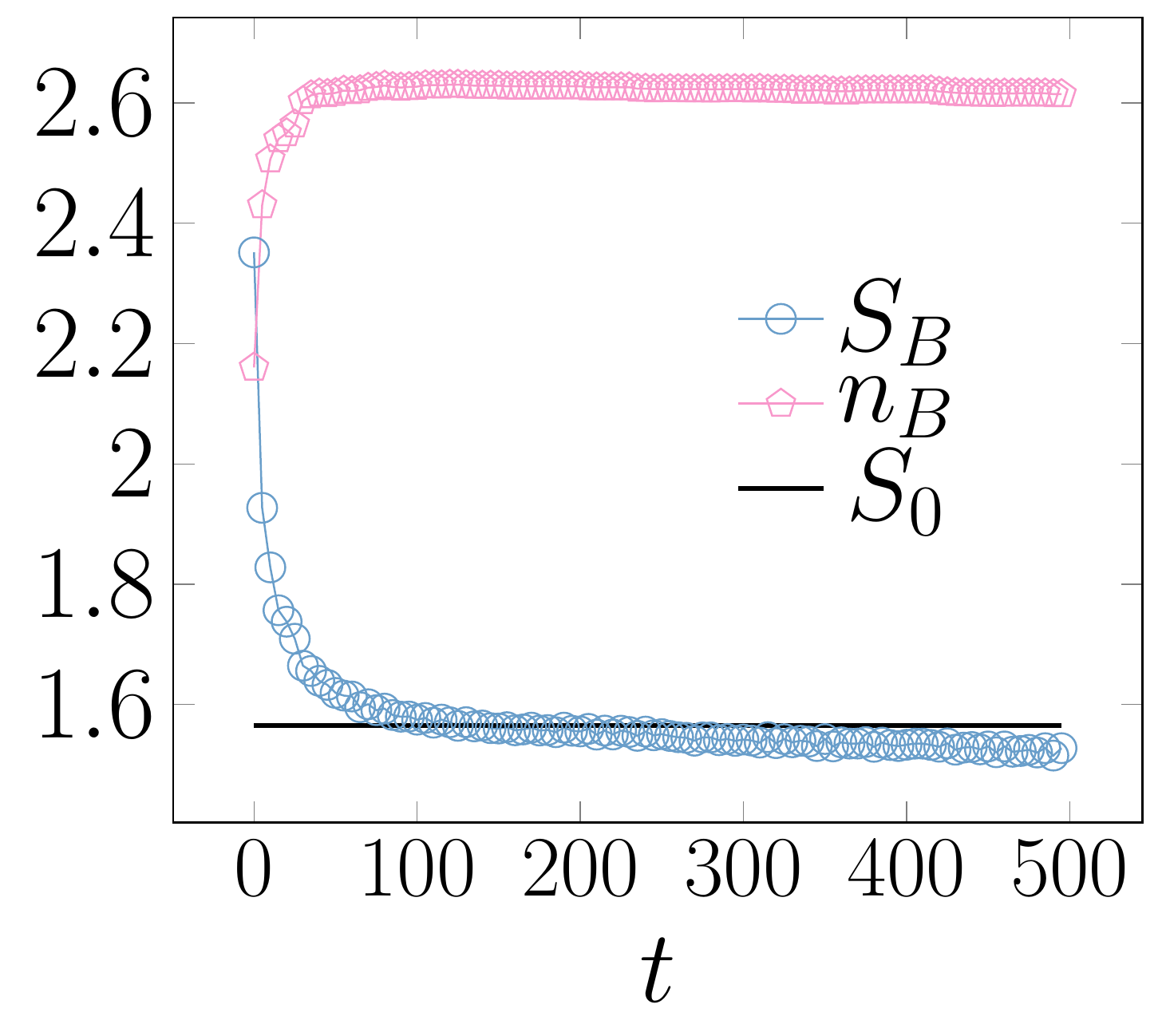}
}
\hspace{-4mm}
\subfigure[]{
\includegraphics[width=4.4cm,height=4.4cm]{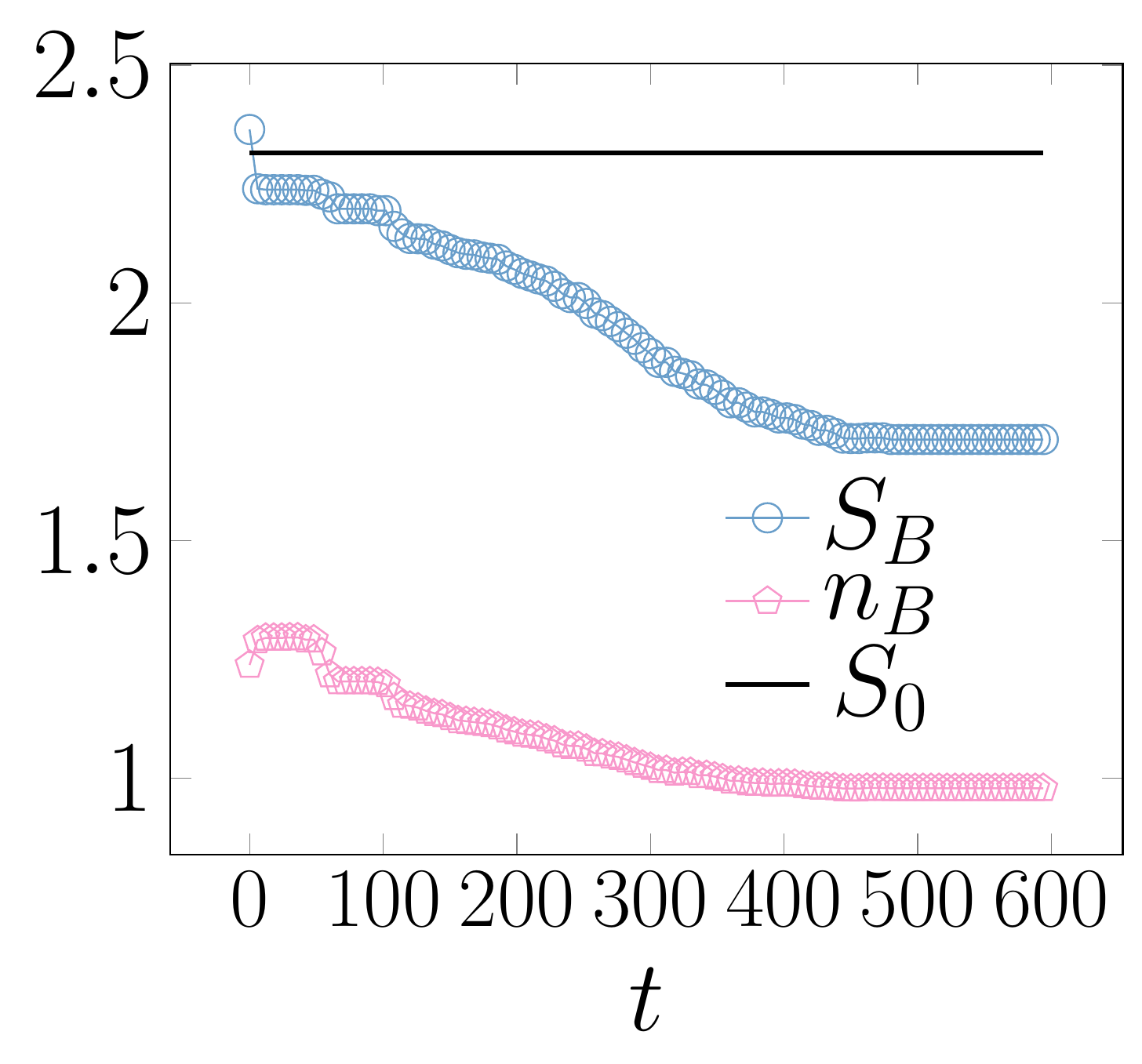}
}
\hspace{-4mm}
\subfigure[]{
\includegraphics[width=4.4cm,height=4.4cm]{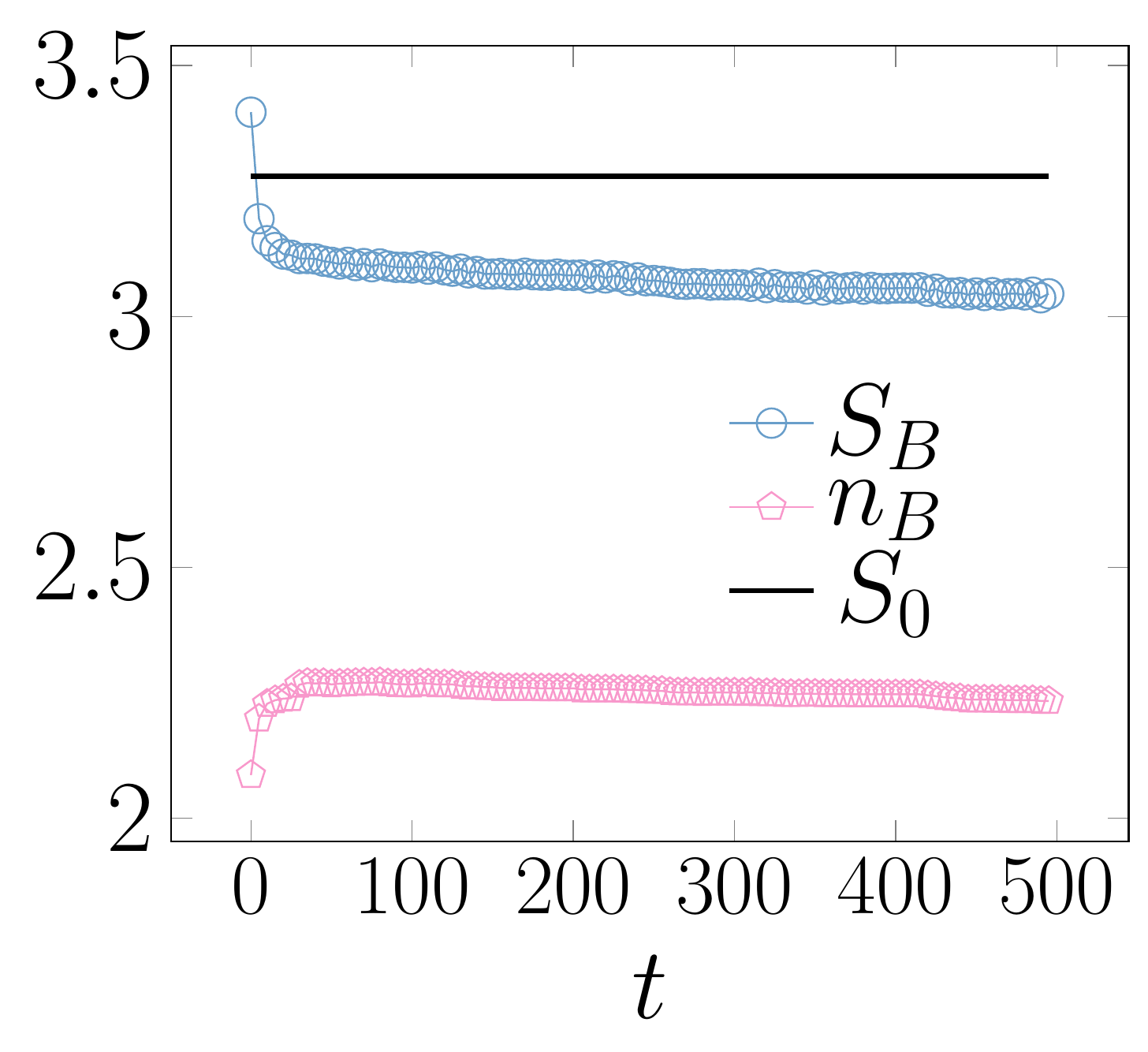}
} 
\caption{
(a)(b) The entropy and Bosons number of system $B$ as the function of $t$. For (a): $L = 5, l_A = 2, N=2$, $\delta t = 0.6$, for (b): $L = 6, l_A = 2, N=3$, $\delta t = 0.5$. The parameter in Bose-Hubbard model is $J = U = 1$, $\beta = 2$. 
(c)(d) Test the control parameters path identified in (a)(b) on the states with $\beta = 1$. 
$S_0$ is the entropy of the total state $\rho_{AB}$. $S_{min}$ is the theoretical lower bound of $S(\rho_B(T))$. The $S_{min}$ for (a) is 0.219, for (b) is 1.067, for (c) is 0.347, for (d) is 1.738.
}
\label{entropy1}
\end{figure*}

In the case where $A$ only has one site, we show that our active quantum distillation groups particles together and decreases the entropy of $\rho_B$ to the entropy of $\rho_{AB}$, which is the lower bound of the entropy of $\rho_B$ under arbitrary unitary transformation. From Theorem~\ref{thm1}, if A only has one site, the optimal $U$ satisfies $\rho_{AB} = U \rho_{AB}^0 U^{\dagger} = \rho_B$, and $\min_U S(\rho_B(U)) = S(\rho_{AB})$.

The entropy of subsystem $B$ as the functions of $t$ is presented in Fig.~\ref{entropy2}(a). The entropy of $B$ decreases to nearly exact the theoretical lower bound in $T = 30$.
The entropy of $A$ and mutual information $I_{AB}$ also decrease under the evolution. The purity of $B$, which is defined as $P_B = \tr \rho_B^2$, increases under the evolution. In Table~\ref{tab:bose}, we list the difference between the final entropy of $\rho_B$ and theoretical lower bound, as well as total evolution time for different size of system. The total evolution time $T$ for $S_{B}$ to converge to the minimum is not scaling with the system size. Actually, the total time $T$ depends on the $\delta t$. Also the entropy of $\rho_B$ does not converge to the lower bound with any $\delta t$. Thus finding the best $\delta t$ needs a fine tuning process. 

In Fig.~\ref{entropy2}(b) we plot the final entropy of subsystem $B$, $S(\rho_B(T))$, as the function of temperature $\beta$, with other parameters fixed. The distillation protocol performs well at all temperatures. When the temperature is high, the theoretical lower bound is very close to the initial entropy of $\rho_B$. However, at low temperature the initial state is more like a pure state, thus subsystem $B$ can be distilled to nearly zero entropy. 

The control parameters path we found based on one state with reverse temperature $\beta$ also works well for other state with $\beta'$, while other parameters $J,U$ are fixed. The test of path for state with $\beta = 1$ on the state with $\beta = 5$ is plotted in Fig.~\ref{entropy2}(c). We find the control parameters path $\Gamma$ using the initial thermal state with $\beta = 1$ and use the same path $\Gamma$ to test the distillation on a thermal state with $\beta = 5$. Under the control parameters path for state with $\beta = 1$, the entropy of $\rho_B$ with $\beta = 5$ also converges to its lower bound with only a little fluctuation. Other stability tests are shown in the Supplemental Material~\cite{SM}.

A key hallmark of quantum distillation is the final state with lower entropy and a greater number of Bosons. Fig.~\ref{entropy2}(d) shows the number of particles in subsystem $B$ as a function of $t$ for different $\beta$. The final state $\rho_B$ groups more Bosons with low entropy. When the initial state tends to be a pure state, the final state $\rho_B$ tends to group all the Bosons in the initial state $\rho_{AB}$. In the case that the subsystem $A$ has only one site, the distillation process mainly eliminates the non-block-diagonal elements in the density matrix of $\rho_{AB}$ such that $\rho_{AB} = \rho_{B}$. Note that
\begin{align} \label{n_b}
    \hat{n}_B = \sum_{i=l_A+1}^{L} \hat{n}_i = \oplus_{k=0}^{N_{AB}} kI_{d_{B,k}\times d_{B,k}},
\end{align} 
where $kI_{d_{B,k}\times d_{B,k}}$ is an identity matrix with dimension $d_{B,k}\times d_{B,k}$. The largest eigenvalues of $\rho_{B}$ are mainly distributes on the bases with largest occupation number $n_B$. The presence of non-block-diagonal elements will diminish the degree of majorization of the diagonal elements of $\rho_{B}$, thus causes $\langle \hat{n}_B \rangle$ to be smaller. So we get the conclusion that distillation process will increase the particle number of subsystem $B$, which is consistent with numerical simulation plotted in Fig.~\ref{entropy2}(d). 

The control parameters path in our protocol is actually an approximately adiabatic evolution. Fig.~\ref{entropy2}(e) shows $\gamma (t)$ as the function of $t$. If we add the number of choices $d$ in $\Gamma$, the parameter $\gamma(t)$ tends to change continuously. The strength of interaction terms in the Hamiltonian can be tuned adiabaticly instead of discretely choosing from $\Gamma$.

From Theorem~\ref{thm1}, we can decrease the entropy of $\rho_B$ lower than that of $\rho_{AB}$ only when $A$ has more than one sites. It is important that in this case we can also groups Bosons together with $S(\rho_B(U)) = S(\rho_{AB})$ by using some control parameters.
However, by fine tuning the control parameters, we can actually find a control path such that $S(\rho_B(U)) < S(\rho^0_{AB})$ with increasing the number of Bosons or only distilling out very few Bosons.
Combining Theorem~\ref{thm1} and Eq.~\eqref{n_b}, we get a corollary that when subsystem $\rho_B$ has lowest entropy, the smallest particle number of $\rho_B$ is $\langle \hat{n}_B \rangle = \sum_{k} k \sum_{b} q_{k,b}$, which is very close to $0$. However we do not want to get a low entropy state with nearly no Bosons. It is important to find a balance between the lower entropy and 
larger number of particles. Fortunately, the unitary transformation constructed by our protocol does not possess strong abilities to completely rearrange all the eigenvalues of $\rho_{AB}$ to get the lowest entropy state.
In Fig.~\ref{entropy1}(a)(b), we plot the entropy of $\rho_{B}$ as the function of $t$, for states with two sites in $A$. The control parameters path is searched using Eq.~\eqref{greedy_algorithm} and Eq.~\eqref{greedy_algorithm_1}. For Bose-Hubbard model with $L=5, l_A = 2$ and $N = 2$, the entropy of $\rho_{B}$ decrease to that half of $\rho_{AB}$ and the Bosons number of $\rho_B$ slightly decrease to to around $N/2$. For Bose-Hubbard model with $L=6, l_A = 2$ and $N = 3$, the entropy of $\rho_{B}$ decrease to slightly lower than that of $\rho_{AB}$ and the Bosons number of $\rho_B$ increase to around $5N/6$.
The test of control parameters path identified in the states with $\beta = 2$ on the states with $\beta = 1$ is plotted in Fig.~\ref{entropy1}(c)(d). 

It is worthy noting that we can also use purity as the target function in Eq.~\eqref{greedy_algorithm}, where the purity is defined as $P(\rho) = \tr(\rho^2)$.  The active quantum distillation protocol based on optimization of purity can also decrease the von Neumann entropy to its minimum. The performance is basically the same as the protocol based on von Neumann entropy. In Supplemental Material~\cite{SM}, we give numerical results about the protocol based on optimization of purity. 

\textit{Summary and outlook.---}
We introduce the active quantum distillation protocol to decrease the von-Neumann entropy without losing many Bosons of the target system by a controlled unitary evolution in a Bose-Hubbard model. Our protocol only needs to control the interaction terms in Hamiltonian between the two subsystem. We derive the analytical lower bound of the target system entropy, which can be naturally generalized to other Bosonic or fermionic models. When one subsystem $A$ only has one site, our protocol can decrease the entropy of the other subsystem $B$ to the entropy of the total state and groups particles together at the end of the evolution, where the entropy of the total state is the lower bound of the subsystem entropy. When one subsystem $A$ has more than one site, our protocol can decrease the entropy of the other subsystem $B$ lower than the total system with increasing the number of Bosons or only distilling out very few Bosons in $B$.
We also find that one control parameters path identified for one set of model parameters can be used in all models with different parameters. The active quantum distillation protocol is actually universal in the whole model parameter space.

In our previous paper~\cite{an2022learning} about the disentanglement using quantum circuit, the theoretical lower bound of one subsystem entropy in spin (qubit) system was given. We show in the Supplemental Material~\cite{SM} that our entropy minimization protocol is also feasible in qubit systems. Actually, we believe that our minimization protocol is universal for a wide range of states of local-Hamiltonians, and will have wide applications in quantum distillation, cooling~\cite{PRXQuantum.4.010332,Rio2011,PhysRevLett.123.170605,PhysRevE.94.032120,PhysRevA.107.032602,PhysRevA.105.022214,PhysRevResearch.2.023120,Lostaglio2015,Skrzypczyk2014} and metrology~\cite{PhysRevLett.96.010401,PhysRevLett.100.220501,PhysRevLett.130.170801,PhysRevX.10.031003,PRXQuantum.3.010202} technologies.

\textit{Acknowledgments.---}
We thank Cheng-Qian Xu for helpful discussions. This work is supported by National Key Research and Development Program of China (Grants No. 2021YFA1402104 and No. 2021YFA0718302) and National Natural Science Foundation of China (Grant No. 12075310).

\bibliographystyle{apsrev4-2}
\bibliography{bib.bib}
\widetext
\clearpage

\vspace{0.3cm}

\section{Proof of theorem 1}
In this section we prove the Theorem 1 in the main text. We will first give one definition and three lemmas in matrix theory~\cite{MatrixAnalysis} for the proof of Theorem 1. 

\begin{definition}\label{definition1}
A $n\times n$ matrix $A = (a_{ij})$ is called doubly stochastic if 
\begin{align}
a_{ij} \geq 0, ~~~\text{for all i, j,} \\
\sum_{i=1}^{n} a_{ij} = 1, ~~~\text{for all j,} \\
\sum_{j=1}^{n} a_{ij} = 1, ~~~\text{for all i.}
\end{align}
\end{definition}

\begin{lemma}\label{lemma1}
(Schur's Theorem) Let A be an Hermitian matrix, let $\text{diag}\{A\}$ denote the vector whose coordinates are the diagonal entries of $A$ and $\lambda\{A\}$ the vector whose coordinates are the eigenvalues of $A$, then $\text{diag}\{A\} \prec \lambda\{A\}$.
\end{lemma}

\begin{lemma}\label{lemma2}

Let $x, y$ be n-dimensional vectors, then the following two conditions are equivalent:

\begin{align}
&\textit{(i)}~ x\prec y.\\ 
&\textit{(ii)}~ \Tr \varphi(x) \leq  \Tr \varphi(y).
\end{align}

\end{lemma}

\begin{lemma}\label{lemma3}
A matrix $A$ is \textit{doubly stochastic} if and only if $Ax\prec x$ for all vectors $x$.
\end{lemma}

\renewcommand{\proofname}{Proof}

\begin{proof}
We begin the proof from the case that A only has one site. In this case $d_B = d_{AB}$, and the trace process actually drop out those non-block-diagonal elements, i.e. $\rho_{B} = \rho_{AB} - \text{\{non-block-diagonal elements\}}$. When $U$ satisfies $\rho_{AB} = \rho_B$, non-block-diagonal elements are zeros, and $\rho_{B} = \rho_{AB}$. If $\rho_{B} \neq \rho_{AB}$, $\rho_{AB}$ has non-zero non-block-diagonal elements. There exists a block-diagonal unitary matrix $U_D$ that can diagonalize $\rho_B$, $U_D\rho_B U_D^{\dagger} = diag\{\lambda(B)\}$, where $diag\{\lambda(B)\}$ represents a diagonal matrix with diagonal elements being the eigenvalues of $B$. Note that $U_D\rho_{AB} U_D^{\dagger} = diag\{\lambda(B)\} + \text{\{non-block-diagonal elements\}}$. By Lemma~\ref{lemma1}, $\lambda(AB) \succ \lambda(B)$. And according to Lemma~\ref{lemma2}, $S(B)\leq S(AB)$.

We next prove general cases. Denote the optimal unitary transformation as $V=V_1 V_{D}$, where $V_{D}$ is an unitary transformation that block-diagonalizes $\rho_{AB}^0$, and $V_1 = \oplus_k V_{1,k}$ is an unitary transformation that diagonalizes $\rho_{AB}^{D} = V_{D} \rho_{AB}^0 V_{D}^{\dagger}$. 
Take the bases of $\mathcal{H}_{A,k}$ as $\{|a_{k}\rangle, 1\leq a_{k}\leq d_{A,k}\}$, and the bases of $\mathcal{H}_{B,k}$ as $\{|b_{k}\rangle, 1\leq b_{k}\leq d_{B,k}\}$.
Let block-diagonal density matrix be $\rho_{AB}^{bd} = \oplus_k \rho_k$, and $\rho_k = \sum_{l=1}^{d_{A,k} d_{B,k}} p_{k,l} |\psi_{k,l}\rangle \langle \psi_{k,l}|$. So $V_{1,k} |\psi_{k,(b-1)d_k^A +a} \rangle = |a,b \rangle _k $.
Then 
\begin{align}
\rho_{B,k} &= \sum_{a,b} p_{k,(b-1)d_{A,k}+a} \Tr_A ( |\psi_{k,(b-1)d_{A,k}+a} \rangle \langle \psi_{k,(b-1)d_{A,k}+a}| ) \nonumber \\
&= \sum_b q_{k,b} |b\rangle \langle b| _k,
\end{align}
where 
\begin{align}\label{qkb_supp}
    q_{k,b} = \sum_{a=1}^{d_{A,k}} p_{k,(b-1)d_{A,k} + a }~~~,
\end{align}
which implies $S(\rho_B (V)) = -\sum_{k}\sum_{b} q_{k,b} \log_2 q_{k,b}$. 

1. If $\rho_{AB} = U \rho_{AB}^0 U^{\dagger}$ is not diagonal but block-diagonal, $ U = U_{D} V_1 V_{D}$, where $U_{D}$ is a block-diagonal unitary matrix. And $U_{D} = \oplus_k U_{D,k} $, where $U_{D,k}$ is an unitary matrix in k'th subspace. The eigen-decomposition of $\rho_{B,k}(U)$ is  
\begin{align}\label{eq_qu1}
    \rho_{B,k}(U) &= \sum_b q_{k,b}(U) |b(U)\rangle \langle b(U)| _k, 
\end{align}
and 
\begin{align}\label{eq_qu2}
    \rho_{B,k} (U)&=\rho_{B,k} (U_{D,k} V_{1,k})\nonumber \\
    &=\sum_{a,b}  p_{k,(b-1)d_k^A+a} \Tr_A ( U_{D,k} |a,b \rangle \langle a,b|_k {U_{D,k}}^{\dagger} ) \nonumber \\
    &=\sum_{m,n} \sum_{a,b}  p_{k,(b-1)d_k^A+a}  \langle m,n^{(U)}|_k U_{D,k} |a,b \rangle
    \langle a,b|_k {U_{D,k}}^{\dagger} |m,n^{(U)}\rangle|n^{(U)}\rangle \langle n^{(U)}|_k \nonumber \\
    &=\sum_{m,n} {p_{(k),m,n}^{(U)}} |n^{(U)}\rangle \langle n^{(U)}|_k,
\end{align}
where 
\begin{align}
    {p_{(k),m,n}^{(U)}} = \sum_{a,b} B_{m,n;a,b}^{(k)}~p_{(k),(b-1)d_{A,k}+a}
\end{align}
with 
\begin{align}
    B_{m,n;a,b}^{(k)} =  \langle m,n^{(U)}|_k U_{D,k} |a,b \rangle_k \langle a,b|_k {U_{D,k}}^{\dagger} |m,n^{(U)}\rangle_k
\end{align}
Combining Eq.~\eqref{eq_qu1} and Eq.~\eqref{eq_qu2}, we arrives at
\begin{align}
    q_{k,b}(U) = \sum_{n} {p_{k,b,n}^{(U)}}~.
\end{align}
Note that $B_{m,n;a,b}^{(k)}\geq 0$, and $\sum_{a,b}B_{m,n;a,b}^{(k)} = \sum_{m,n}B_{m,n;a,b}^{(k)}= 1$, namely, $B$ is a doubly stochastic matrix. According to Lemma~\ref{lemma3}, we obtain 
\begin{align}
    {p_{k}^{(U)}} \prec p_k~.
\end{align}
Now we denote $p_{k,\downarrow}^{(U)}$ as $p^{(U)}_{k}$  in decreasing order. Similarly, we introduce $q_{k,\downarrow,b}^{(U)} = \sum_{a} {p_{\downarrow , k,(b-1)d_k^A + a }^{(U)}}$. According to the definition of majorization, 
\begin{align}
    q_{\textbf{}\downarrow}^{(U)} & \prec q_{k}~, \\
    q^{(U)}_{k} & \prec q_{k,\downarrow}^{(U)}~,
\end{align}
which implies that $q^{(U)}_{k} \prec q_{k}$. According to Lemma~\ref{lemma2},
\begin{align}
    S(\rho_A (V)) \leq S(\rho_A (U))).
\end{align}
This completes first part of the proof.

2. If $\rho_{AB}$ is not block-diagonal, let $\rho^{D}_{AB}$ is the block-diagonal part of $\rho_{AB}$. There exists an $U^{D}$, such that $U^{D} \rho^{D}_{AB} U^{D\dagger} = diag\{\lambda(\rho_{AB}^{D})\}$. Note that $U^{D} \rho_{AB} U^{D\dagger} = diag\{\lambda(\rho_{AB}^{D})\} + \text{\{non-block-diagonal elements\}}$. By Schur-Horn theorem, $\lambda(\rho_{AB}) \succ \lambda(\rho_{AB}^{D})$. Then we repeat the proof in part 1, and we finish our proof of part 2.

3. Our protocol in Theorem is actually as follows: divide the eigenvalues of $\rho_{AB}^0$ into each subspace, and the larger eigenvalues are in the subspace with larger $d_A$. When $\rho_{AB}$ is diagonal, each $q_{k,b}$ is the sum of $d_{A,k}$ eigenvalues of $\rho_{k}$. So we want the larger $d_{A,k}$ is, the larger $p_{k,l}$ are. Then by definition of majorization, when $\rho_{AB}$ is diagonal, any order that differs from the above protocol will result a set $\{q'\}$, such that $q'\prec q$, where $q$ is the set in Eq.~\eqref{qkb_supp}. This completes the proof of our theorem.

\end{proof}

\section{Theoretical lower bound of one subsystem entropy in particle number non-conservation Bose-Hubbard model}

In this section we analyse the case of particle number non-conservation system, in which we assume the particle number is upper bounded by $N_{max}$, i.e., the system is classically superposed by mixed states with different occupation number bases.
The density matrix of such system is block diagonal with $N_{max}$ sectors, $\rho_{AB}^{N_{max}} = \oplus_{n_{AB}=0}^{N_{max}} \rho_{AB}^{n_{AB}}$, where $\rho_{AB}^{n_{AB}}$ is the sector that only has bases with occupation number $n_{AB}$. 

In order to give the analytical expression for this kind of system, we are going to re-divide the sectors of the density matrix according to the occupation number of $B$. For a sector with occupation number $n_{AB}$, the block diagonolization is,
\begin{align}
   \rho_{AB}^{n_{AB}}=& \mathop{\oplus}\limits_{n_A+n_B=N_{AB}} \Tilde{\rho}_{n_A,n_B}\nonumber \\
                                            &+ \text{\{non-block-diagonal elements\}}.
\end{align}
Note that the minimization of entropy is realized when the non-block-diagonal elements are $0$'s. So we suppose the density matrix is fully block-diagonal,
\begin{align}
     \rho_{AB}^{n_{AB}} = \oplus_{n_B=0}^{N_{max}} \rho_{n_B}^{n_{AB}-n_B},
\end{align}
and the density matrix of the total system $\rho_{AB}^{N_{max}}$ becomes
\begin{align}
     \rho_{AB}^{N_{max}} = \oplus_{n_{AB}=0}^{N_{max}} \oplus_{n_B=0}^{N_{max}} \rho_{n_B}^{n_{AB}-n_B}.
\end{align}
Change the order of direct sum, 
\begin{align}\label{nB_blocks}
     \rho_{AB}^{N_{max}} &=  \oplus_{n_B=0}^{N_{max}} \oplus_{n_{AB}=0}^{N_{max}} \rho_{n_B}^{n_{AB}-n_B} \nonumber \\
     &= \oplus_{n_B=0}^{N_{max}} ~\rho_{n_B}.
\end{align}
$\rho_{AB}^{N_{max}}$ is block diagonalized according to the occupation number of $B$.


Note that $\rho_{n_B}$ has tensor product structure, i.e., $\mathcal{H}_{n_B} = \mathcal{H}_{A,n_B} \otimes \mathcal{H}_{B,n_B}$. Let the dimension of $\mathcal{H}_{A,n_B}$ is $d_{A,n_B}$. Now we rearrange the order of $\rho_{n_B}$ such that $d_{A,n_B}$ is in decreasing order:
\begin{align}\label{nB_blocks_1}
     \rho_{AB}^{N_{max}} &= \oplus_{k=0}^{N_{max}} ~\rho_{k},
\end{align}
where $d_{A,k}\geq d_{A,k+1}$. And we denote the total dimension of $AB$ of each $\rho_{k}$ as $d_k$. 

Note that Eq.~\eqref{nB_blocks_1} has the same form as in the main text for number conservation case. Then we can directly use Theorem 1 for the particle number non-conservation system to get the lower bound for the entropy of system $B$ under a unitary transformation.

\section{Additional Numerical Details}

\begin{figure*}[htbp]
\subfigure[]{
\includegraphics[width=4.5cm,height=4.5cm]{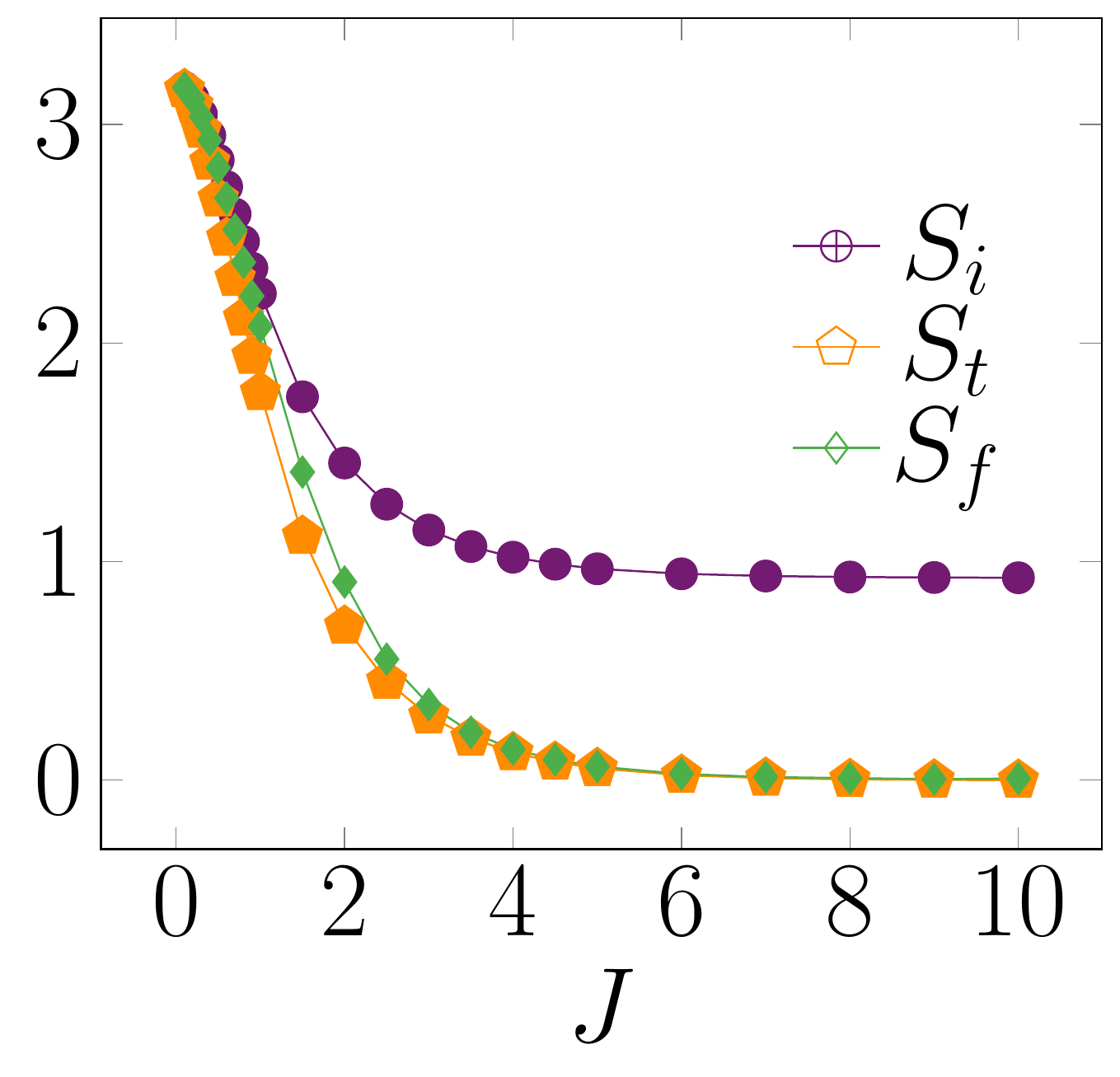}
}
\hspace{-2mm}
\subfigure[]{
\includegraphics[width=4.5cm,height=4.5cm]{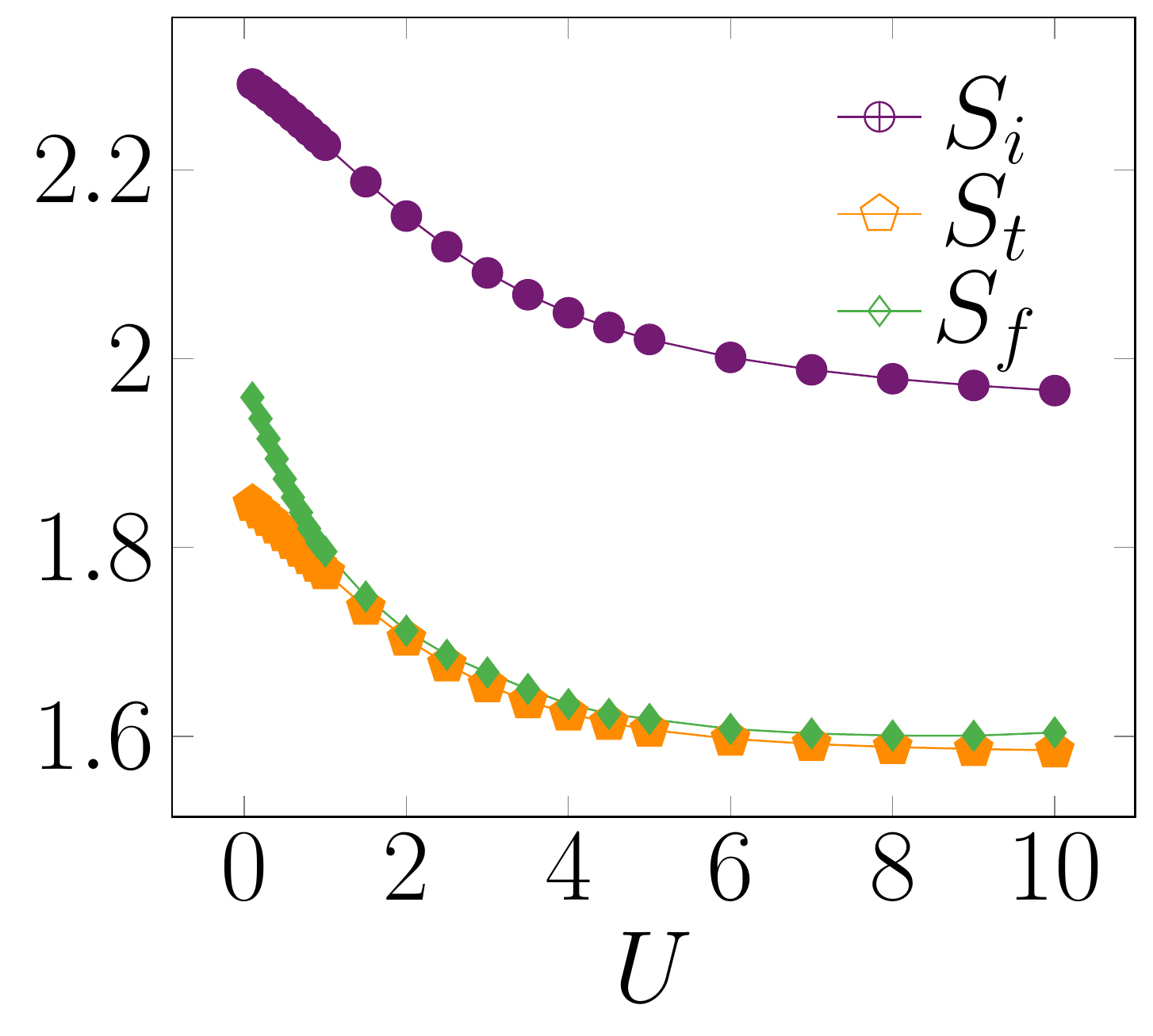}
}
\hspace{-2mm}
\subfigure[]{
\includegraphics[width=4.5cm,height=4.5cm]{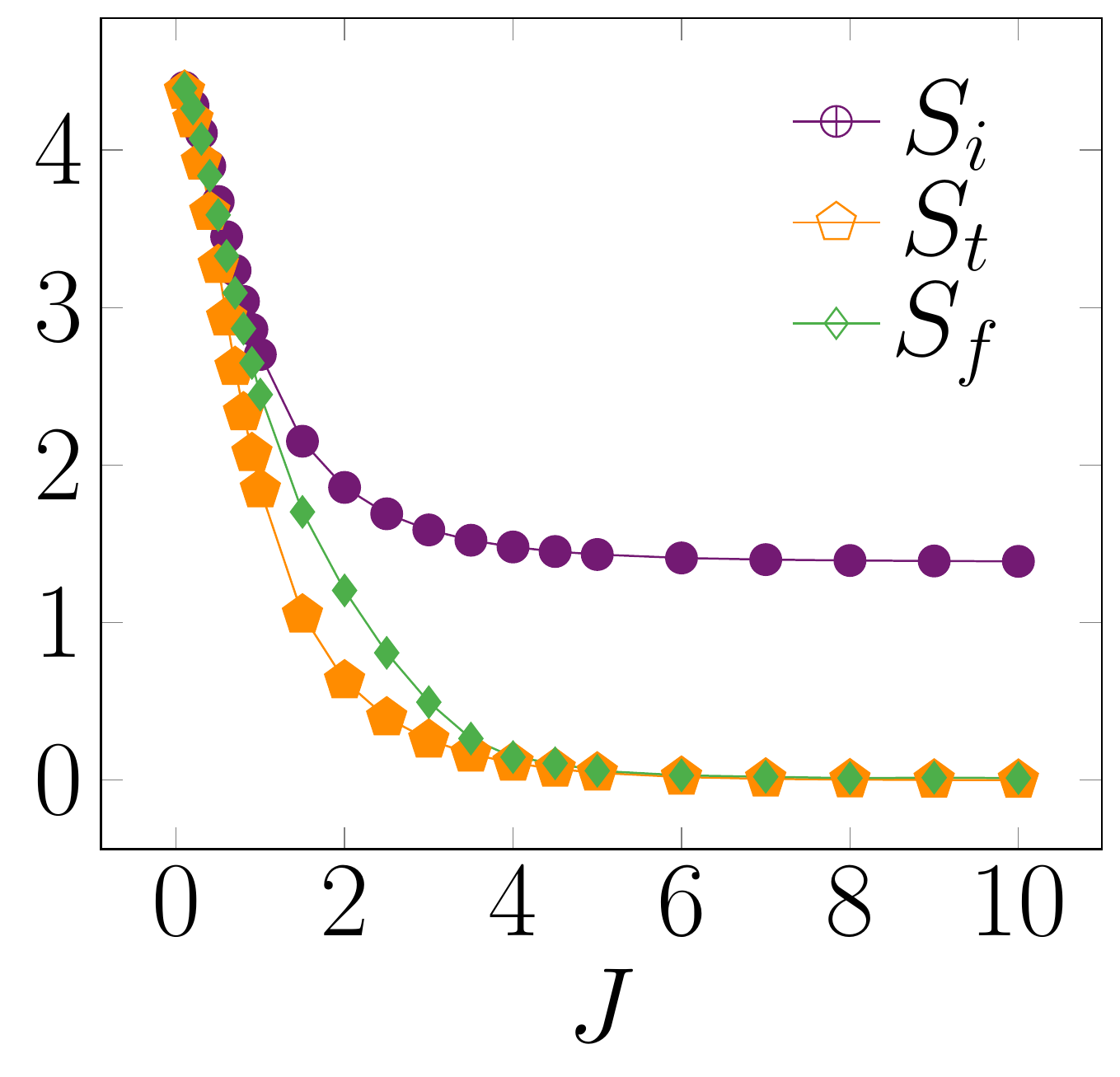}
}
\hspace{-2mm}
\subfigure[]{
\includegraphics[width=4.5cm,height=4.5cm]{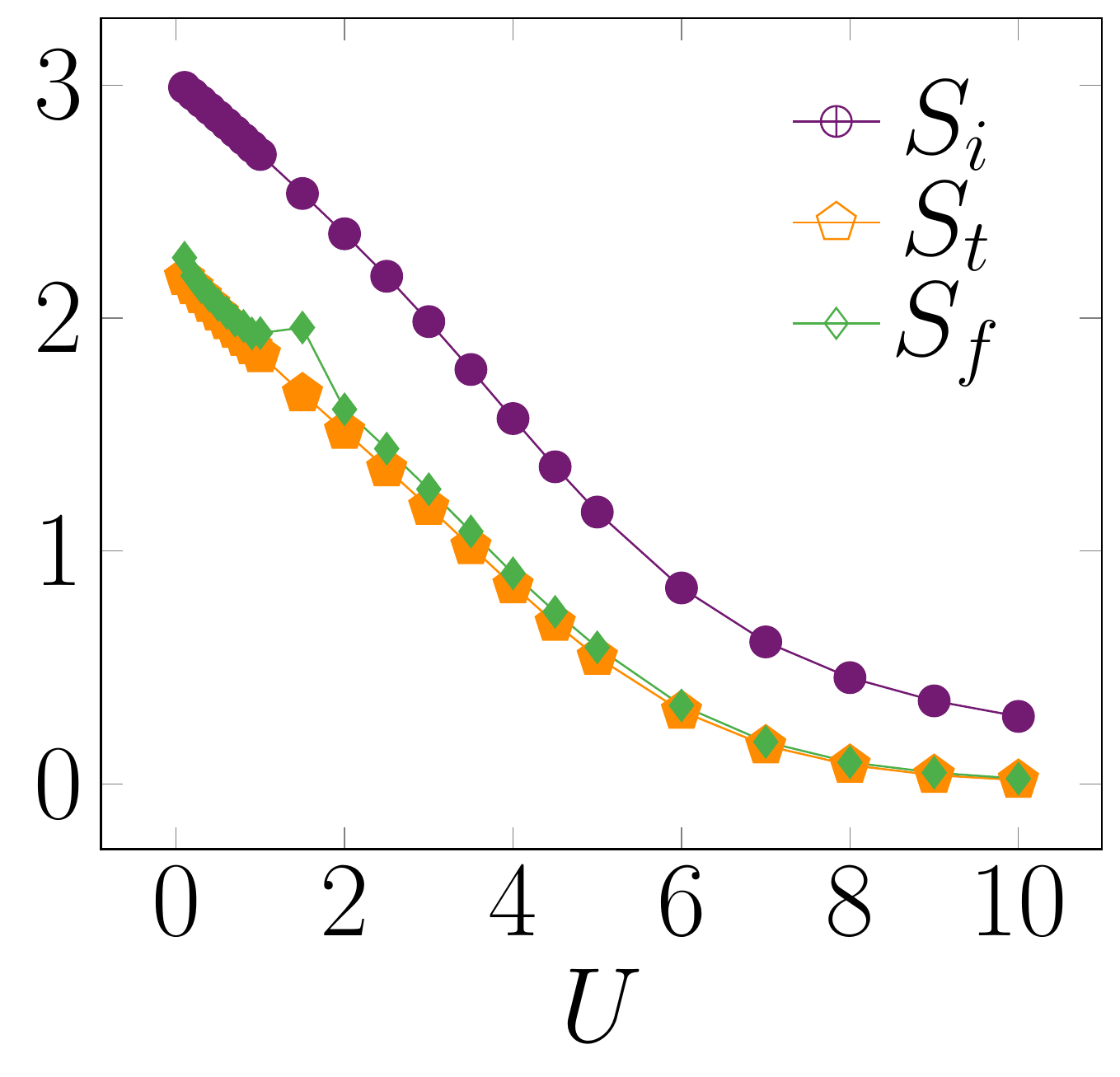}
}
\hspace{-2mm}
\subfigure[]{
\includegraphics[width=4.5cm,height=4.5cm]{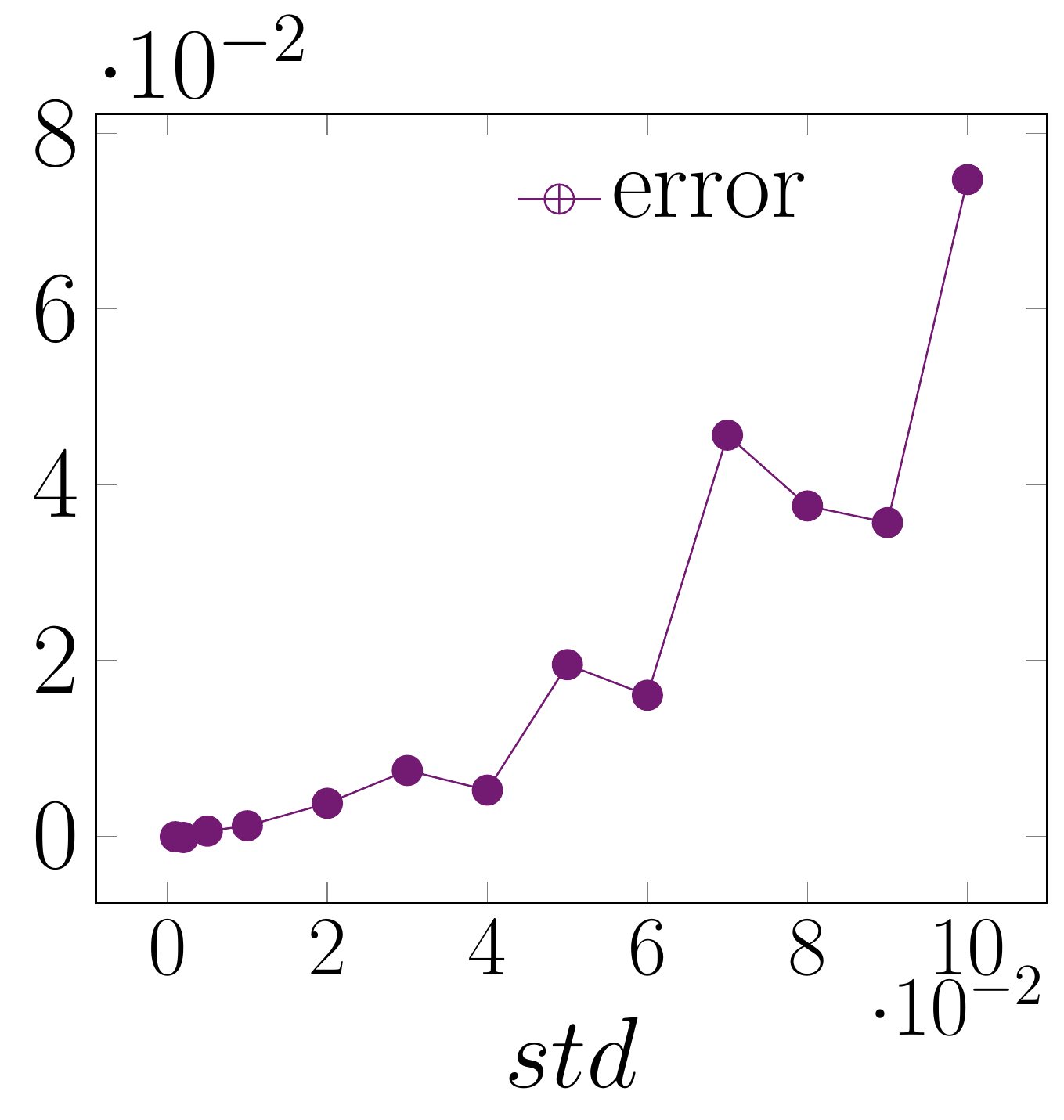}
}
\hspace{-2mm}
\subfigure[]{
\includegraphics[width=4.5cm,height=4.5cm]{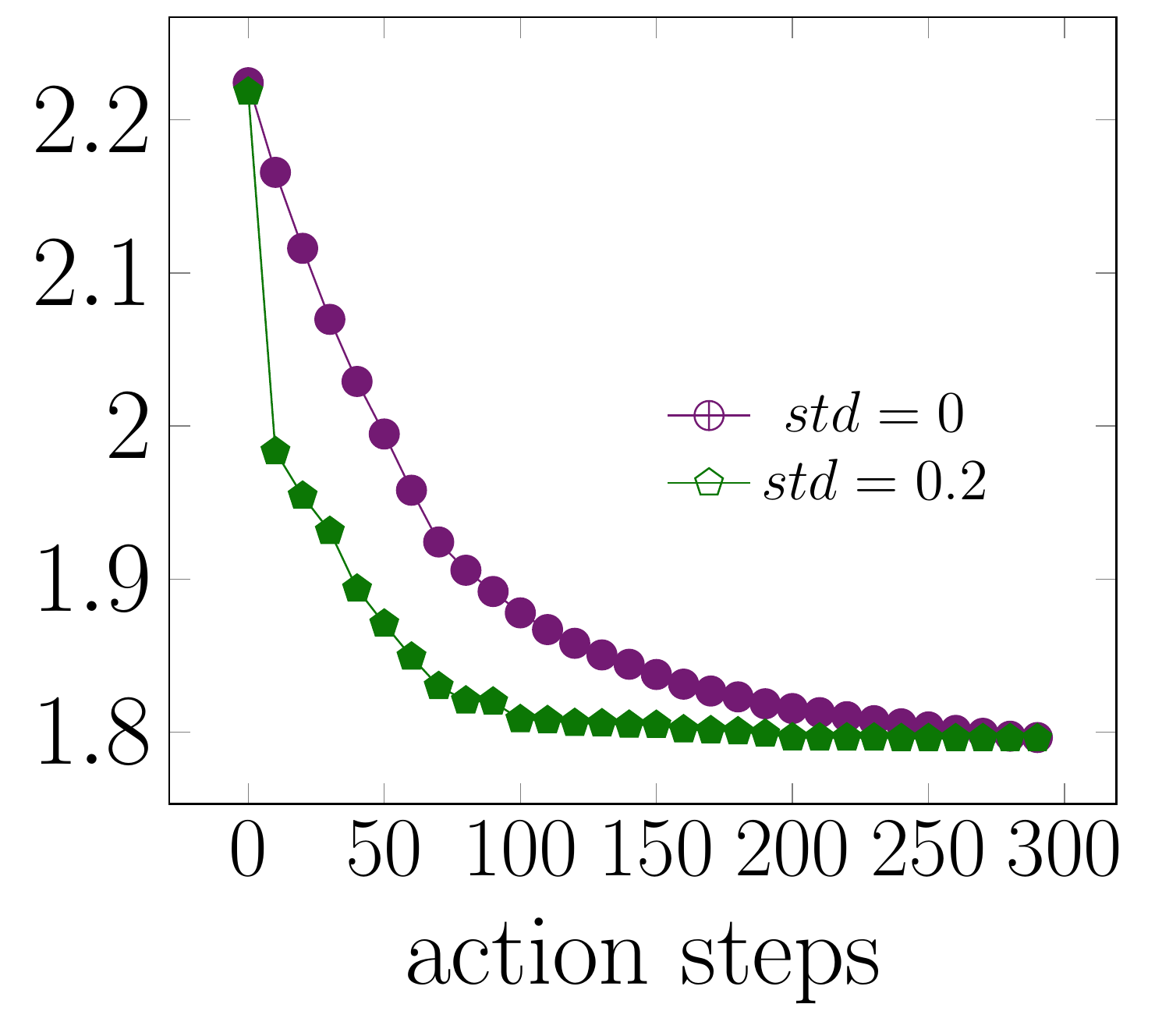}
}

\caption{
(a)(b)The final entropy of subsystem $B$, $S(\rho_B(T))$ as the function of $J$ and $U$, on the Bose-Hubbard model with $L = 4$, $l_{A} = 1$, $N=2$ and $\beta = 1$. $S_i$ is the initial entropy of $\rho_{B}$, and $S_t$ is the theoretical lower bound of $S(\rho_B(T))$. (a) The final entropy of subsystem $B$ with different initial state parameter $J$, with $U = 1$. (b) The final entropy of subsystem $B$ with different initial state parameter $U$, with $J = 1$.
\\
(c)(d)The final entropy of subsystem $B$, $S(\rho_B(T))$ as the function of $J$ and $U$, on the Bose-Hubbard model with $L = 4$, $l_{A} = 1$, $N=4$ and $\beta = 1$. (c) The final entropy of subsystem $B$ with different initial state parameter $J$, with $U = 1$. (d) The final entropy of subsystem $B$ with different initial state parameter $U$, with $J = 1$.
\\
(e) The relative error of the final entropy with imperfect timekeeping evolution as functions of variance of the timekeeping tick $\sigma$ for states with $L = 4, N=2$, $dt = 0.1$.
(f) The comparison between method with and without random $dt$.
}
\label{entropy_app}
\end{figure*}

In this section we give additional numerical details about the active quantum distillation protocol.

In Fig.~\ref{entropy_app}(a)-(d), we plot the final entropy of subsystem $B$, $S(\rho_B(T))$ as the function of $J$ and $U$ on Bose-Hubbard models with different parameters. When the hopping strength is large, Bosons move faster than the small hopping amplitude, therefore larger hopping strength leads to a shorter evolution time. For a model with $N=2$ and $L=4$, when the onsite repulsion $U$ is small, the model is in free Bosons limit $U\to 0$, the evolution time becomes longer. Strong onsite repulsion makes the model to be in a hard-core Bosons limit $U\to \infty$, the  the evolution time becomes longer. However for a model with unit filling~($N=4$ and $L=4$, $N/L=1$), large $U$ implies fast minimization.

We also numerically show that our protocol is robust under the impact of imperfect timekeeping. The impact of imperfect timekeeping on quantum control was shown in~\cite{PhysRevLett.131.160204}: for cooling a qubit with protocol based on a swap gate, timekeeping error only impacts the rate of cooling but not the achievable temperature. In a time duration $\tau$, the imperfect timekeeping means the evolution of an initial state $\rho$ is given as
\begin{align}
    \rho ' = \int_{-\infty}^{\infty} dt \frac{e^ { \frac{(t-\tau)^2}{-2\sigma ^2} } }{\sqrt{2\pi\sigma}} e^{-iHt} \rho e^{iHt}, 
\end{align}
where the $\sigma$ is the variance of the timekeeping tick distribution. Here we numerically show that our protocol is stable against the imperfect timekeeping from the aspects of entropy minimization. Specifically, we first find the optimal control sequence with perfect timekeeping, then we numerically test our optimal control sequence with a Gaussian-error imperfect timekeeping.

In Fig.~\ref{entropy_app}(e) we show the relative error of the final entropy with imperfect timekeeping evolution as functions of variance of the timekeeping tick $\sigma$ with $dt$ is $0.1$. Our protocol retains valid even when the variance of the timekeeping tick distribution $\sigma$ equals to $dt$.

To furthermore shorten the total evolution time, the randomness can be used in this protocol. The time step $dt$ can be chosen as a random Gaussian variable with mean $\langle dt\rangle = dt_0$ and variance $\sigma$. And we use the above method to find control parameters in each random time step $dt_j$. This random time step method does not promise that every path we searched will have a shorter evolution time, however, we find that there is a very high probability that we will search out a better path than the method without randomness. The landscape of the entropy is very complex, taking $dt$ as a random variable to some extends increase the probability to find a better path with shorter total evolution time. In Fig.~\ref{entropy_app}(f) we show the comparison between the method with random time period and without random time period.

\section{Active quantum distillation based on optimization of purity}
In this section we give the numerical results of the active quantum distillation protocol based on optimization of purity. We change the target function in the main text 
\begin{align}\label{greedy_algorithm_app}
    \gamma (t_j) = \underset{\gamma_{j}^{(k)}} {\operatorname{argmin}}~S(\rho_{B,j}^{{(k)}}),
\end{align}
to the purity of the subsystem $B$, 
\begin{align}\label{greedy_algorithm_purity}
    \gamma (t_j) = \underset{\gamma_{j}^{(k)}} {\operatorname{argmax}}~P(\rho_{B,j}^{{(k)}}),
\end{align}
where the purity of the subsystem $B$ is defined as
\begin{align}
    P(\rho_B) = \tr(\rho_B^2).
\end{align}

\begin{figure}[htbp]
\subfigure[]{ 
\includegraphics[width=4cm,height=4cm]{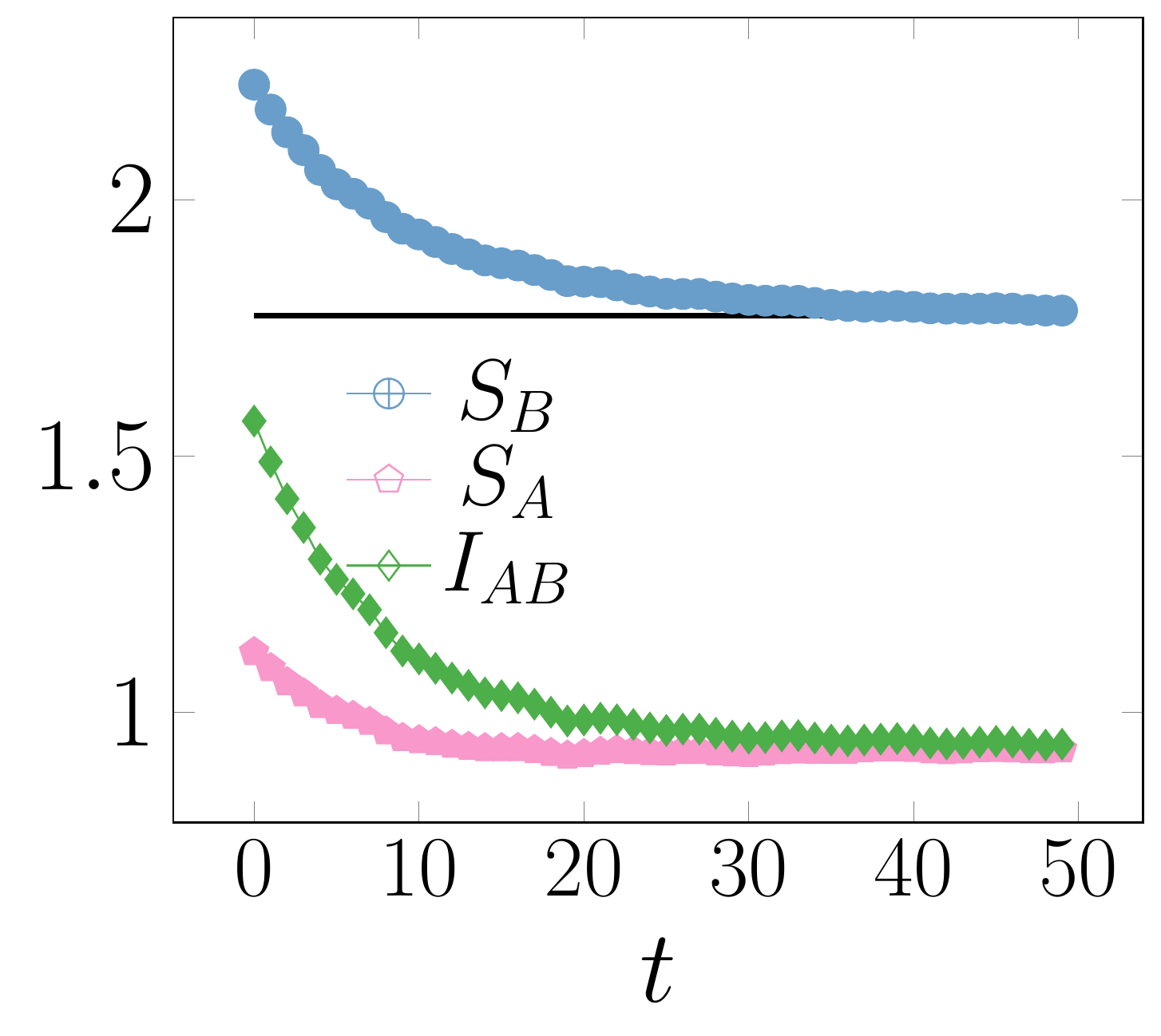}
}
\makebox[0pt][l]{\hspace{-1.9cm}\raisebox{1.2cm}{\includegraphics[width=1.5cm,height=1.5cm]{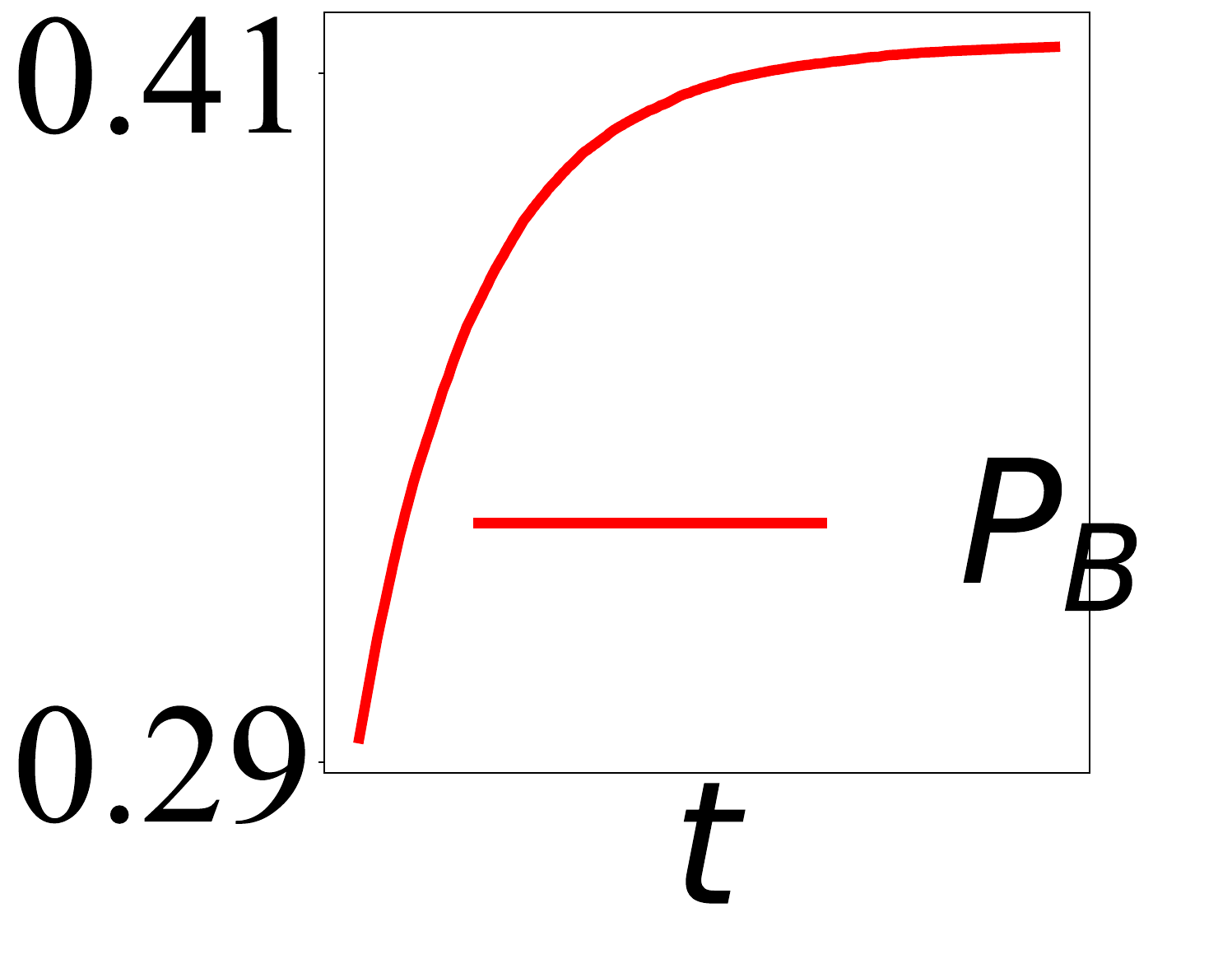}}}
\hspace{-4mm}
\subfigure[]{
\includegraphics[width=4cm,height=4cm]{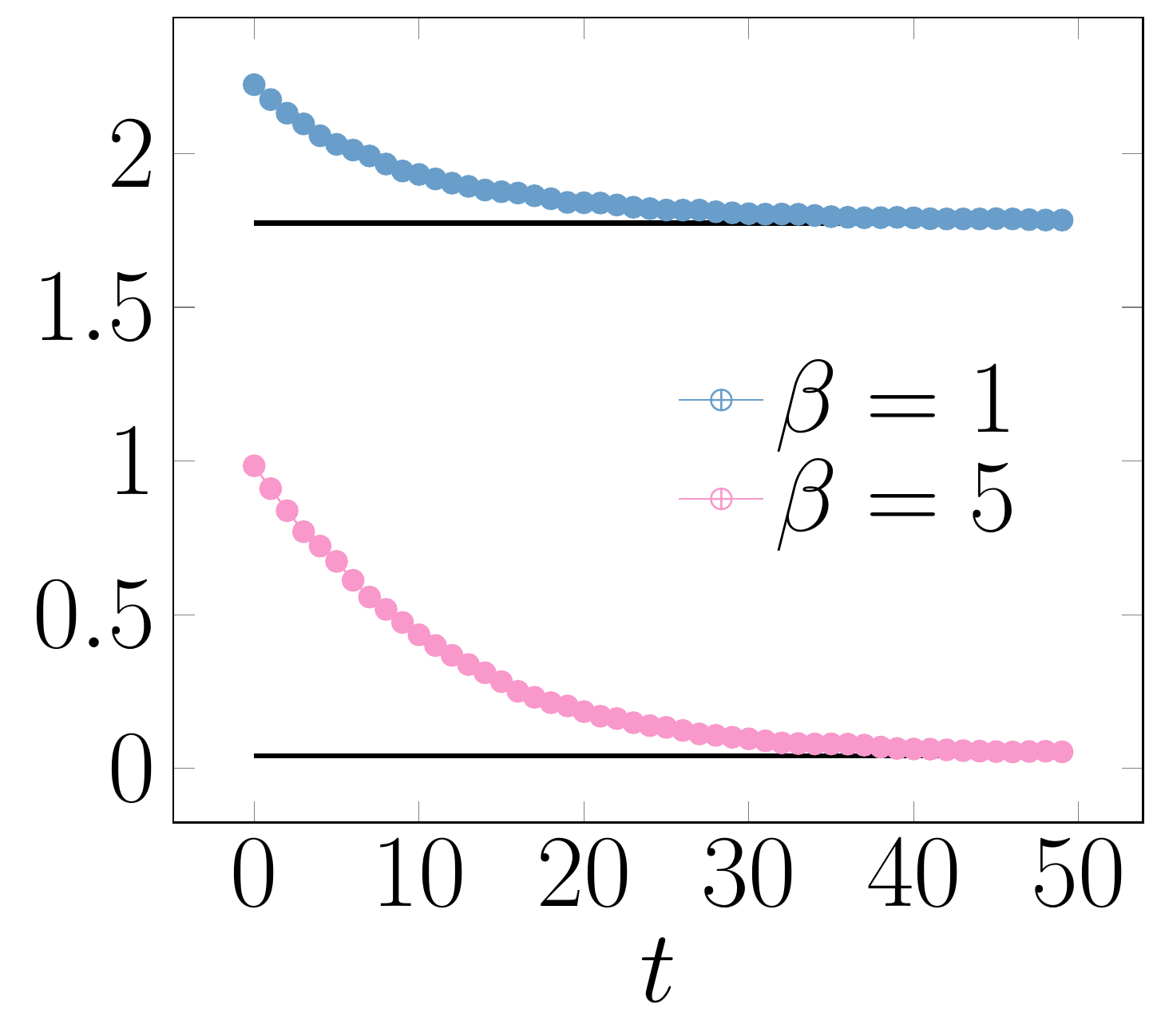}
}
\hspace{-4mm}
\subfigure[]{
\includegraphics[width=4cm,height=4cm]{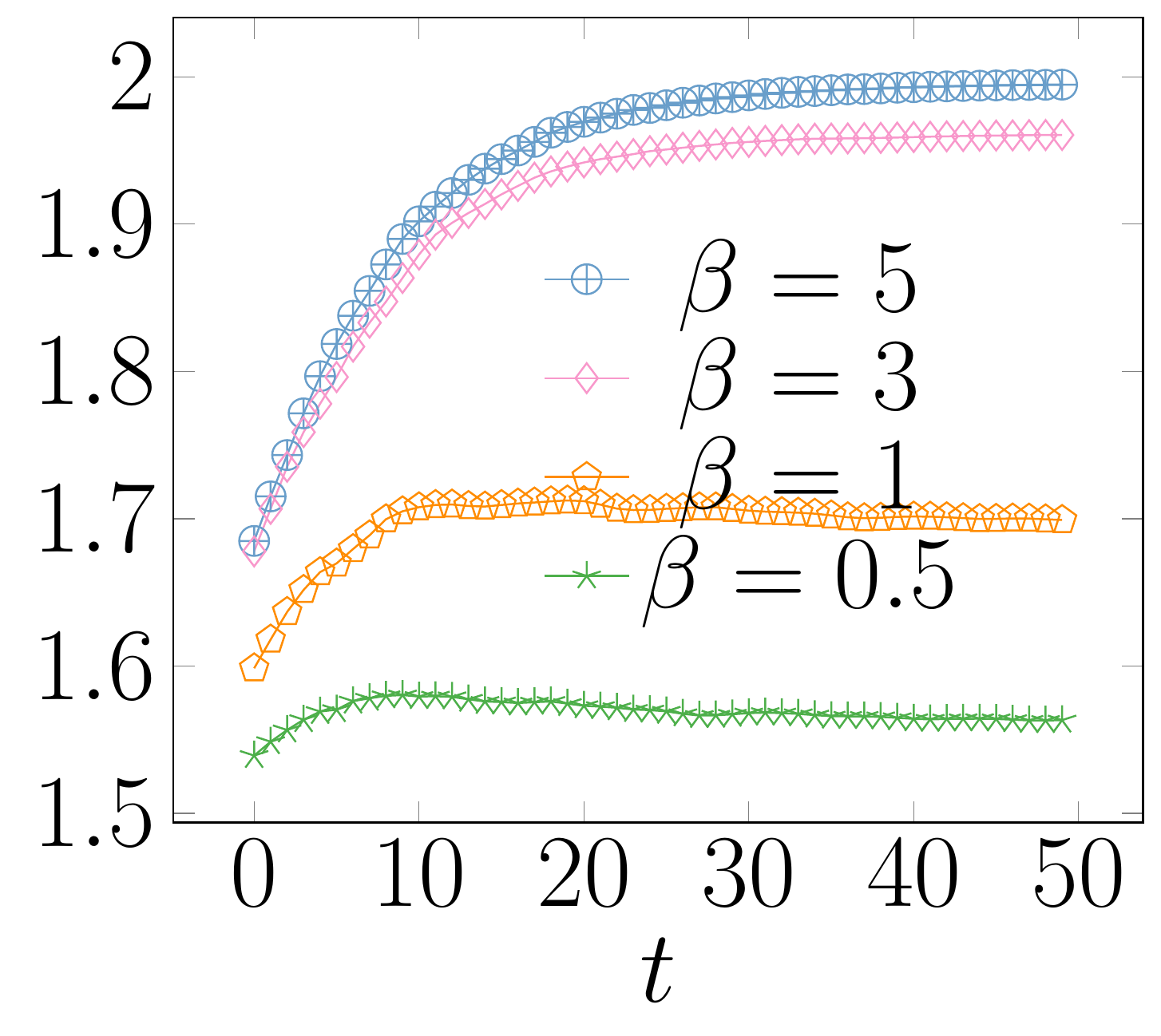}
}
\hspace{-4mm}
\subfigure[]{
\includegraphics[width=4cm,height=4cm]{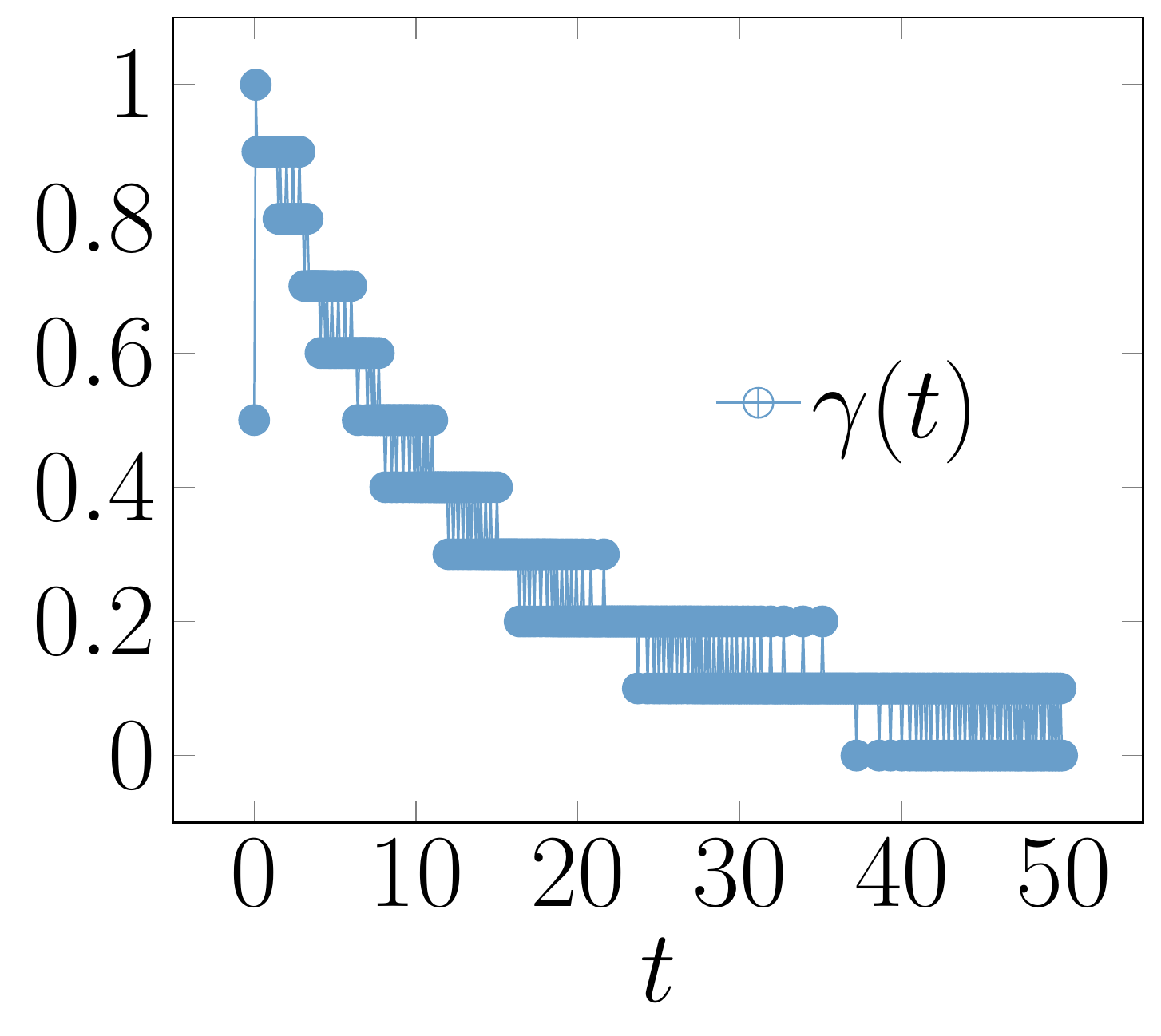}
}
\hspace{-4mm}
\subfigure[]{
\includegraphics[width=4cm,height=4cm]{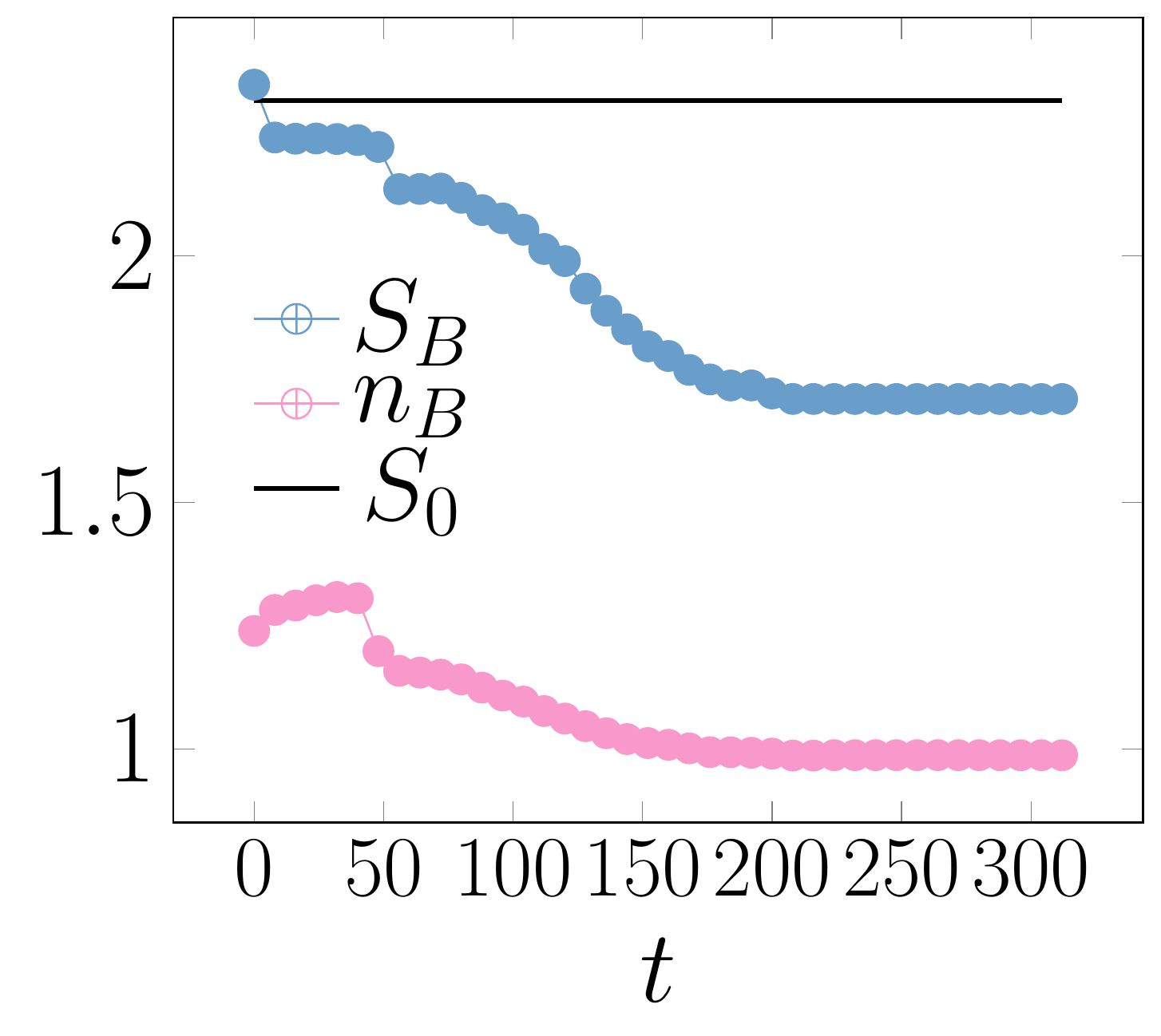}
}
\makebox[0pt][l]{\hspace{-2cm}\raisebox{1.15cm}{\includegraphics[width=1.25cm,height=1.25cm]{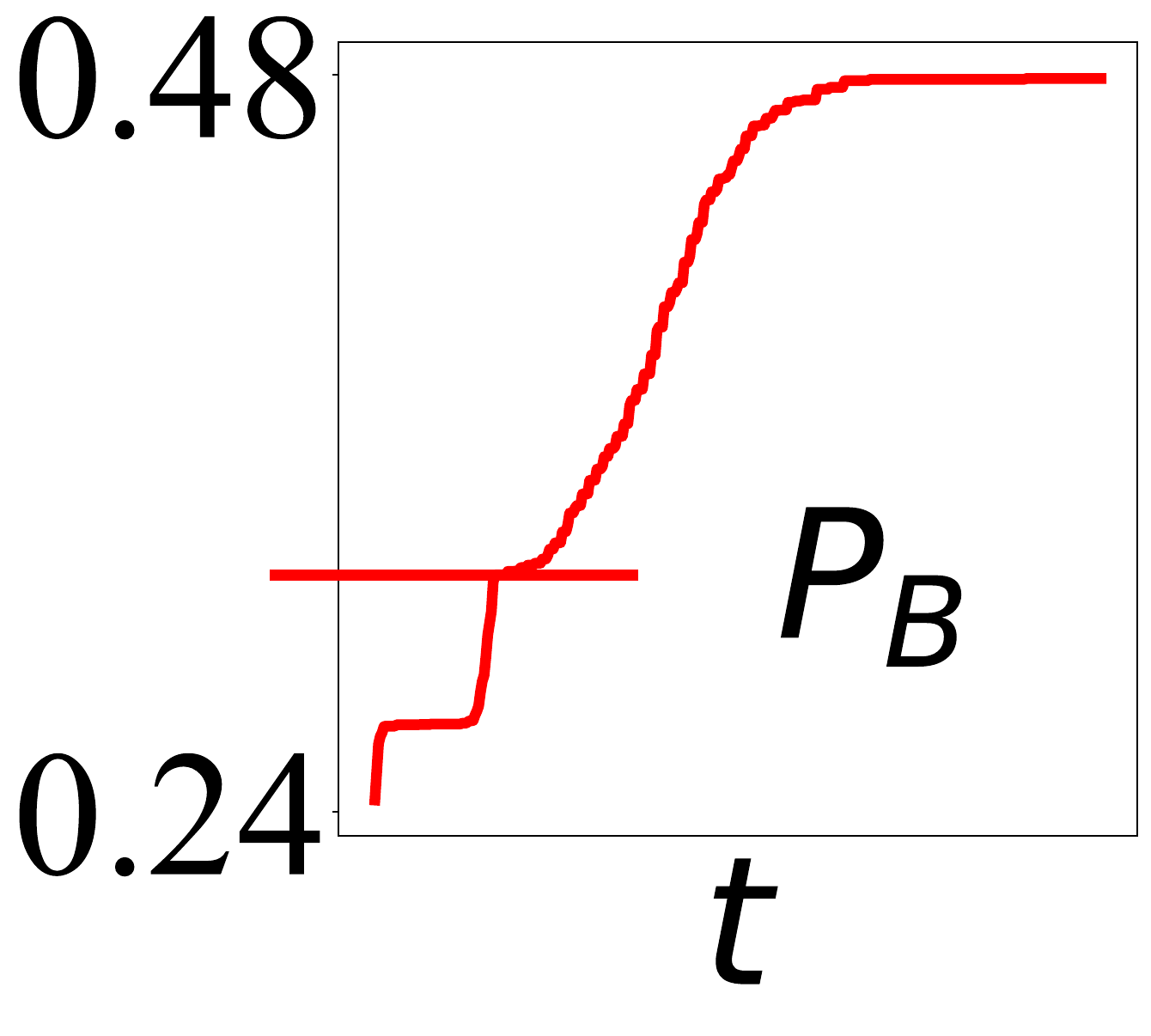}}}
\hspace{-4mm}
\subfigure[]{ 
\includegraphics[width=4cm,height=4cm]{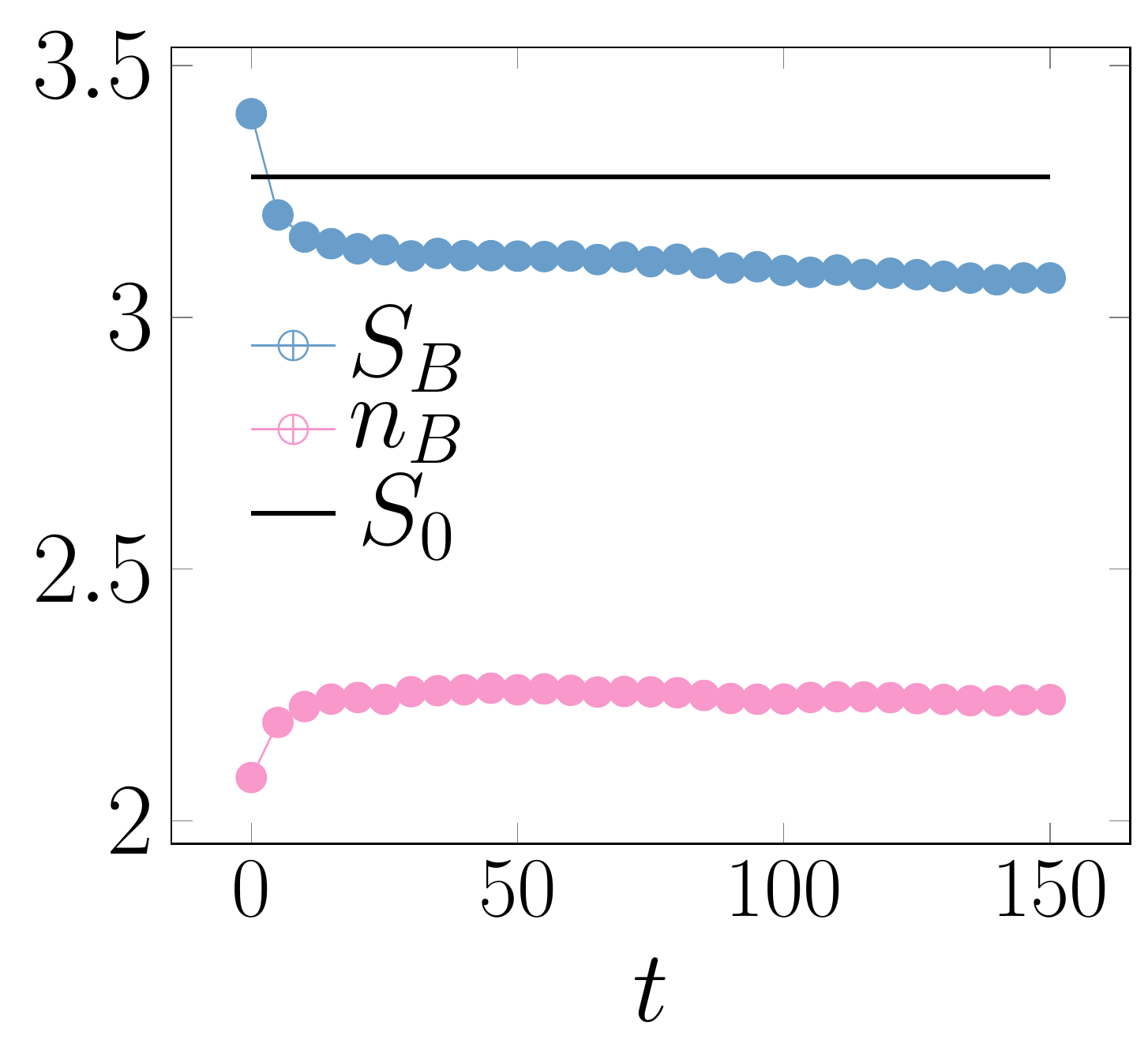}
}
\makebox[0pt][l]{\hspace{-2.1cm}\raisebox{1.4cm}{\includegraphics[width=1.5cm,height=1.5cm]{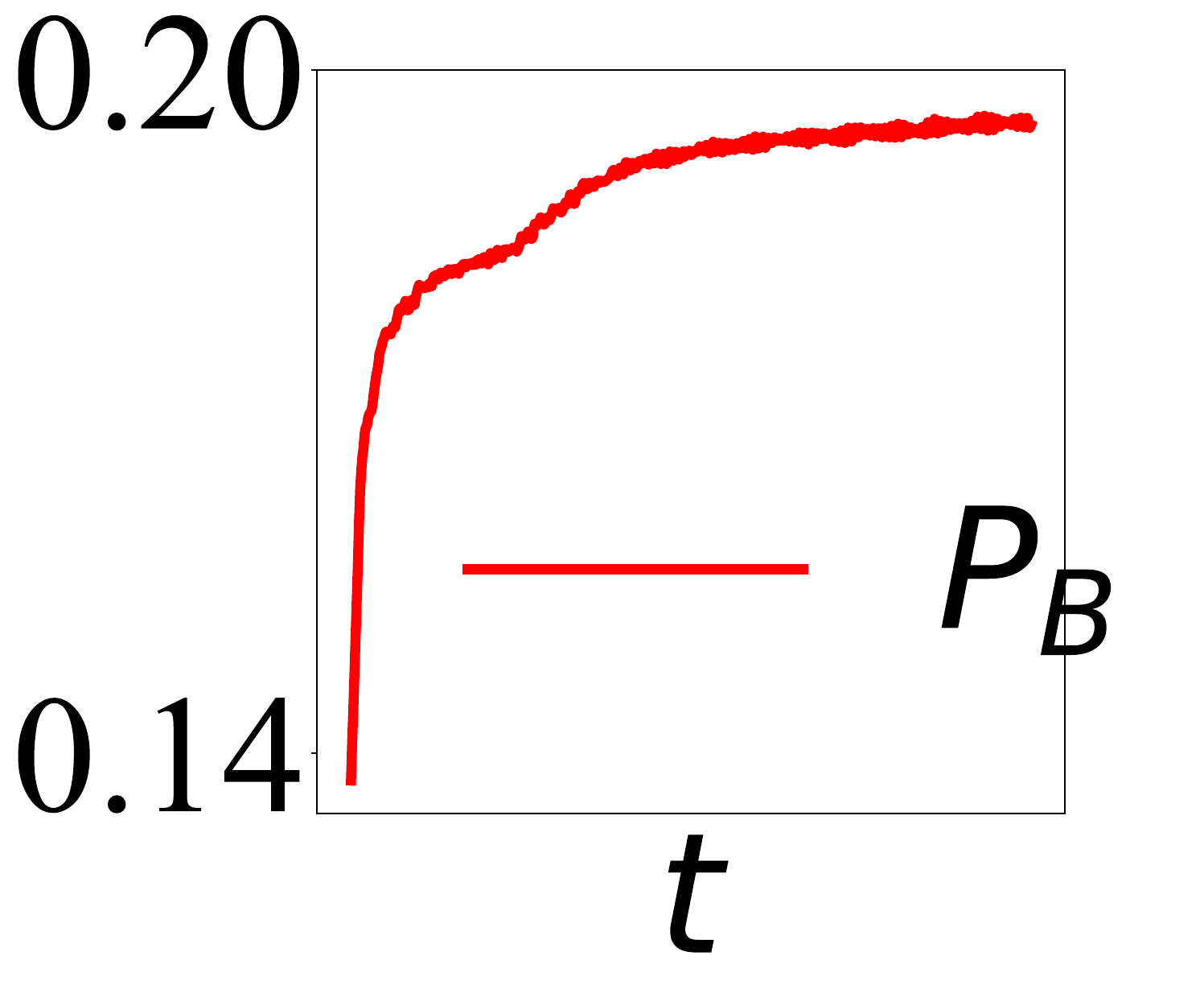}}}
\hspace{-4mm}
\subfigure[]{
\includegraphics[width=4cm,height=4cm]{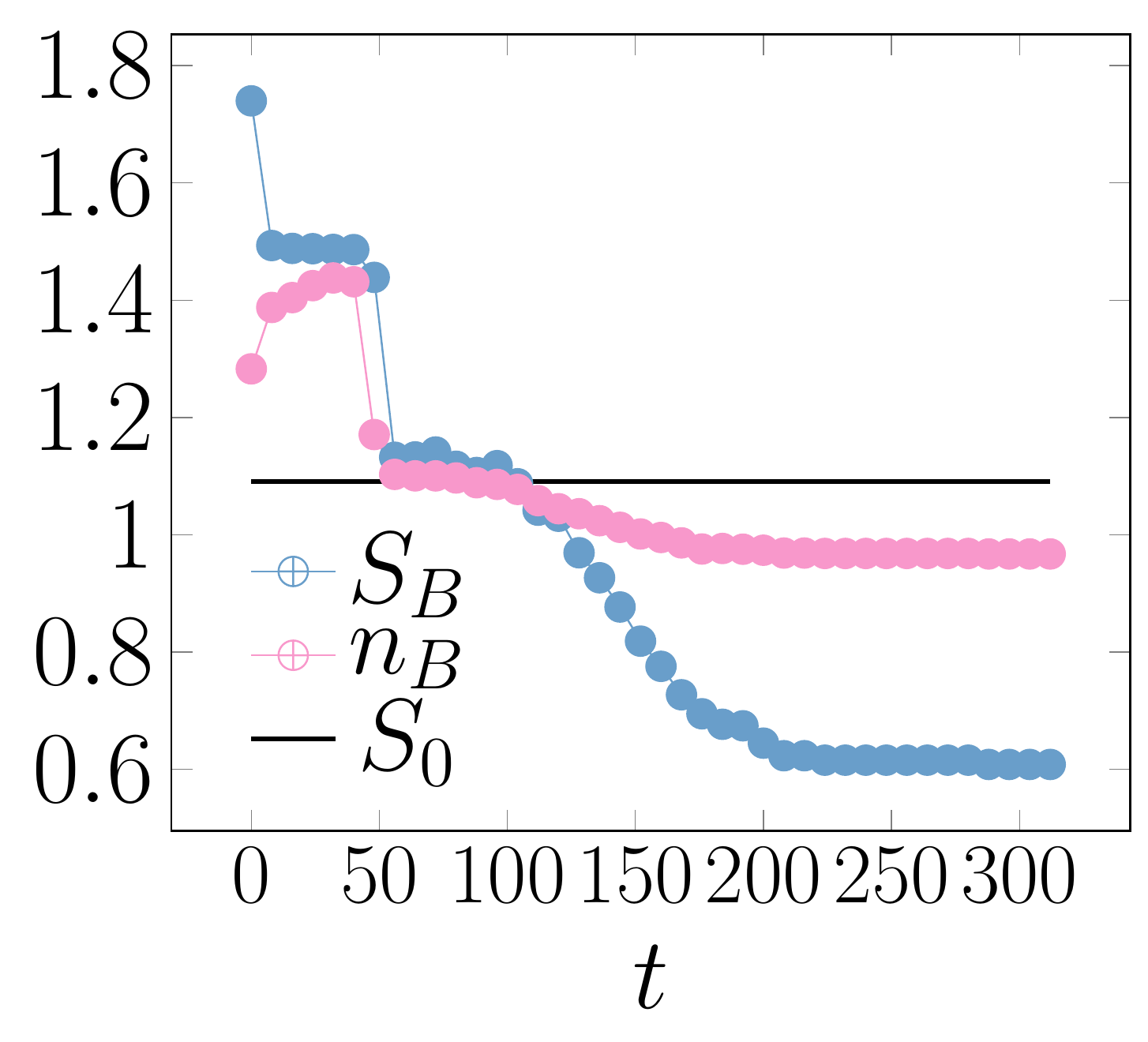}
}
\hspace{-4mm}
\subfigure[]{ 
\includegraphics[width=4cm,height=4cm]{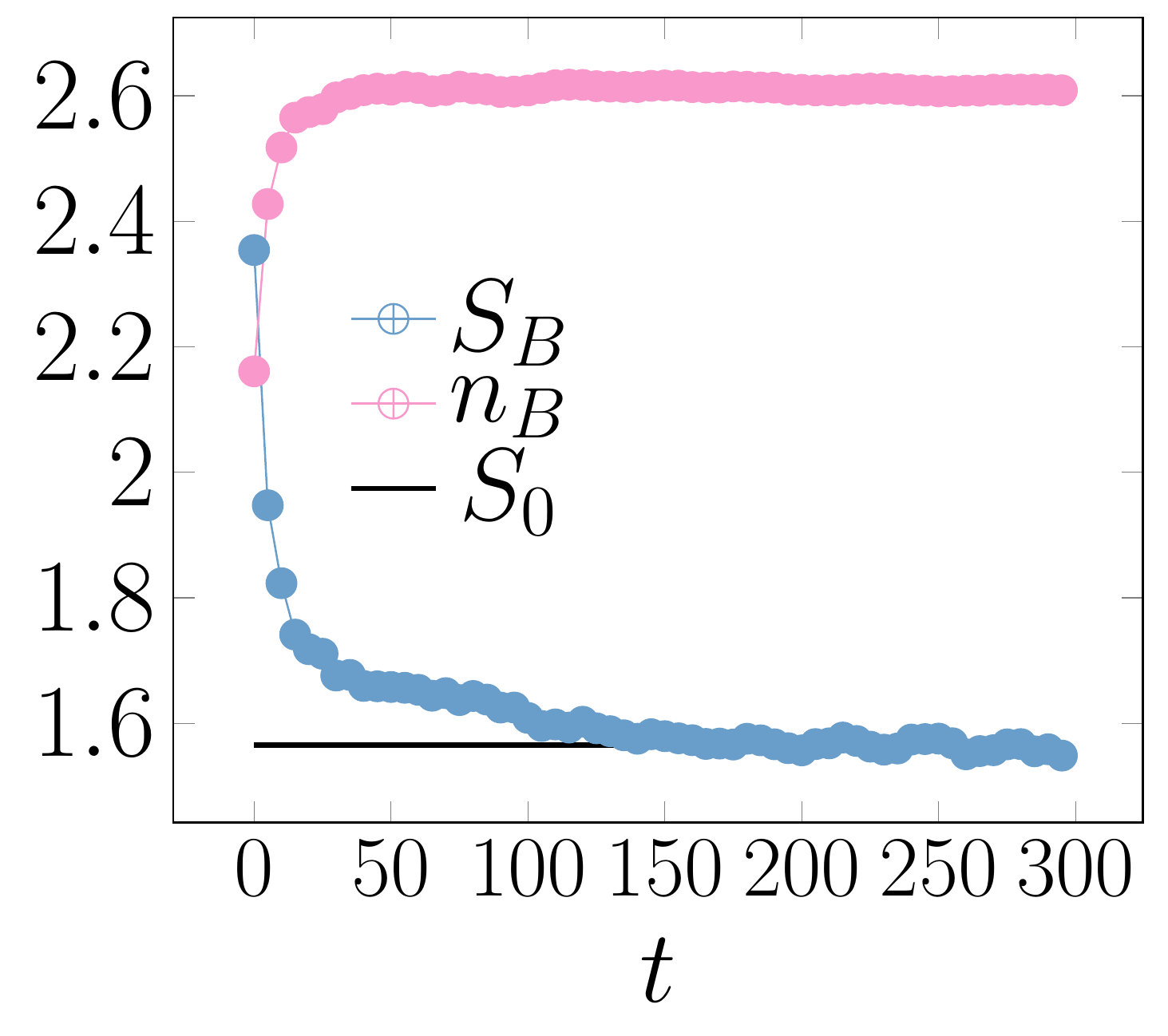}
}
\caption{
The numerical results using the purity as the target function. 
(a) The entropy of subsystem $B$ as the function of time $t$. $\beta =1$, $\delta t = 0.1$.
Inset: the purity of subsystem $B$ as the function of time $t$.
(b) The entropy of subsystem $B$ as the function of time $t$ for $\beta =1$ and $\beta =5$, the control parameters are identified on model with $\beta =1$ and are tested on model with $\beta =5$.
(c) The number of Bosons $\langle \hat{n}_B \rangle$ in subsystem $B$ as the function of time $t$ for different $\beta$.
(d) The choice of $\gamma$ as the function of $t$. The path of the $\gamma(t)$ tends to be an adiabatic evolution.
(e)(f) The entropy and Bosons number of system $B$ as the function of $t$. For (e): $L = 5, l_A = 2, N=2$, $\delta t = 0.8$, for (f): $L = 6, l_A = 2, N=3$, $\delta t = 0.5$, where $\beta = 2$. The parameter in Bose-Hubbard model is $J = U = 1$, $\beta = 1$. Inset: the purity of subsystem $B$ as the function of time $t$.
(g)(h) Test the control parameters path identified in (e)(f) on the states with $\beta = 2$. 
}
\label{entropy_purity}
\end{figure}

The numerical results based on optimization of purity are given in Fig.~\ref{entropy_purity}. In the case where $A$ has one site, using the purity as the target function is almost the same as the using the von Neumann entropy. In the case where $A$ has more than one site, a slight different time step $\delta t$ will be used when using the purity as the target function.

\section{Entropy minimization protocol for Transverse field Ising model}

\begin{figure}[htbp]
\subfigure[]{ 
\includegraphics[width=4cm,height=4cm]{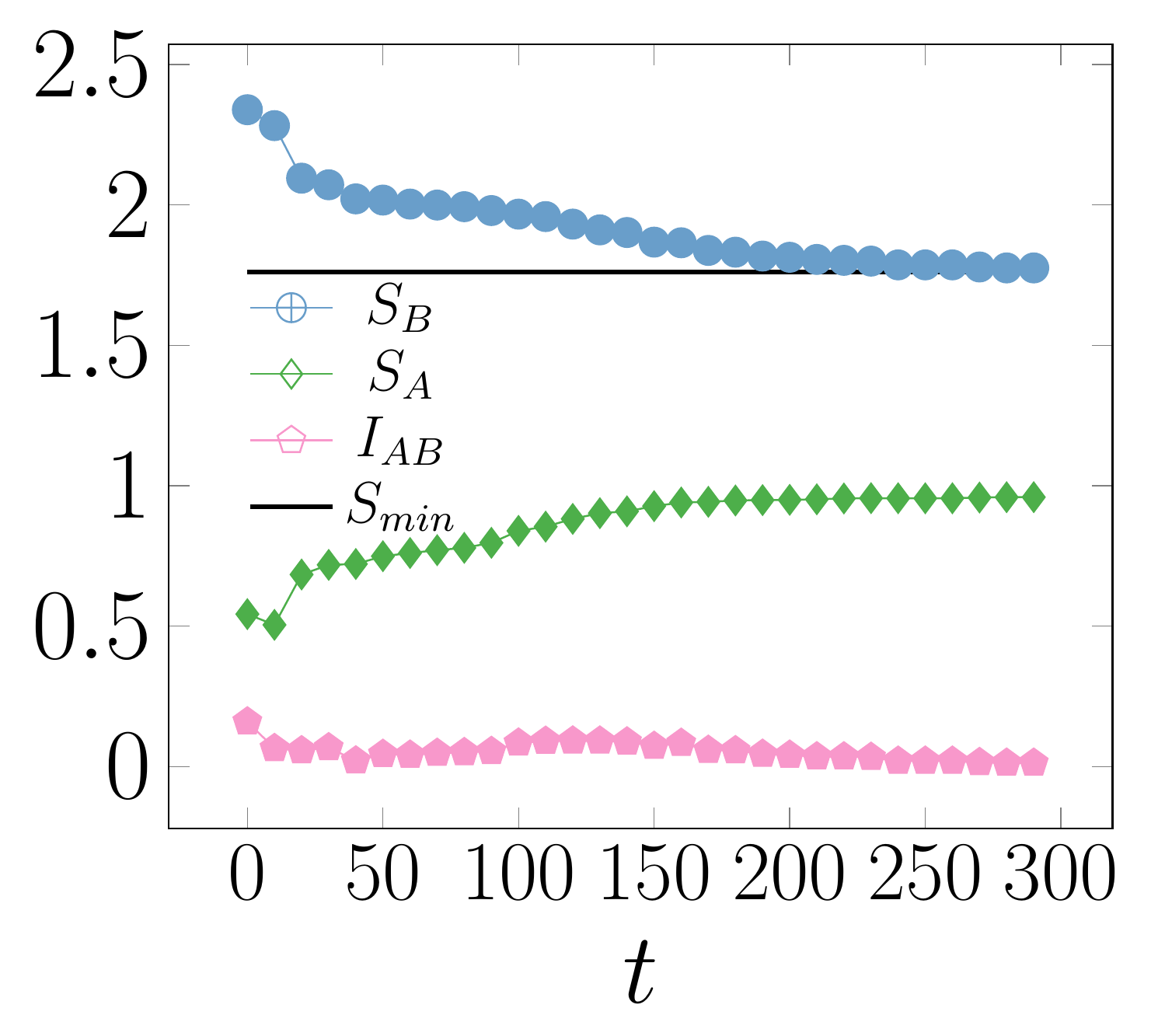}
}
\hspace{-4mm}
\subfigure[]{
\includegraphics[width=4cm,height=4cm]{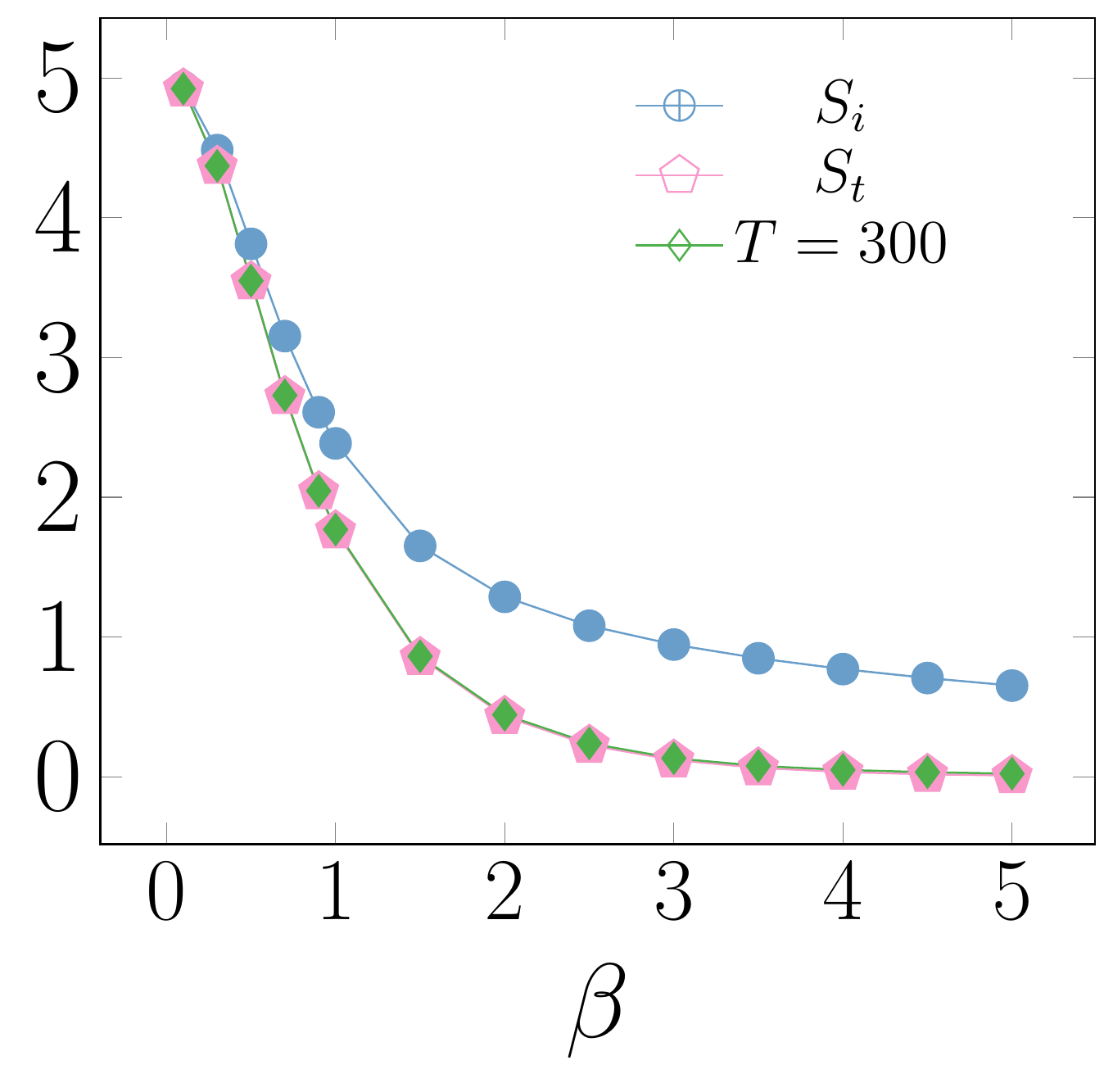}
}
\hspace{-4mm}
\subfigure[]{
\includegraphics[width=4cm,height=4cm]{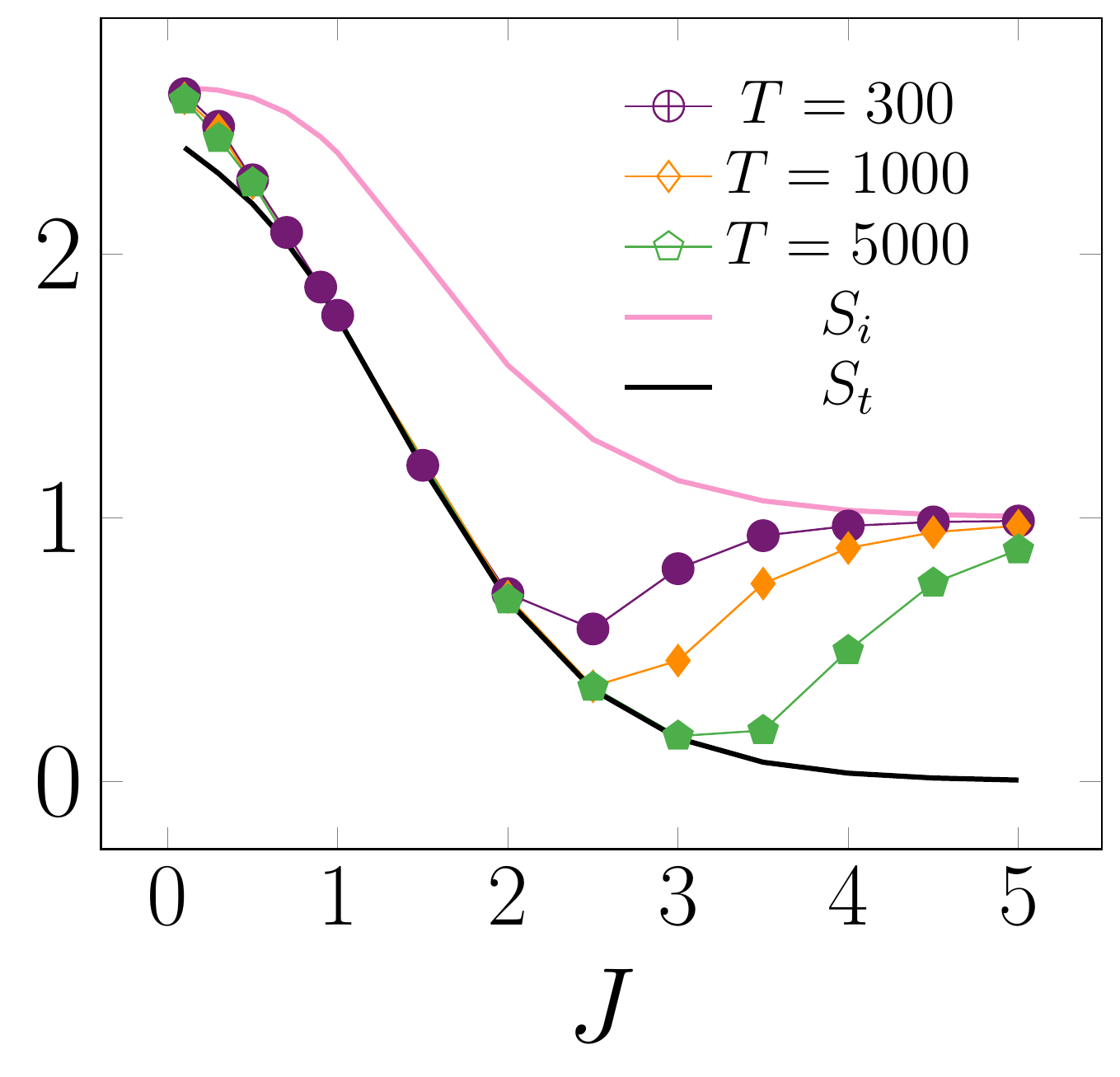}
}
\subfigure[]{ 
\includegraphics[width=4cm,height=4cm]{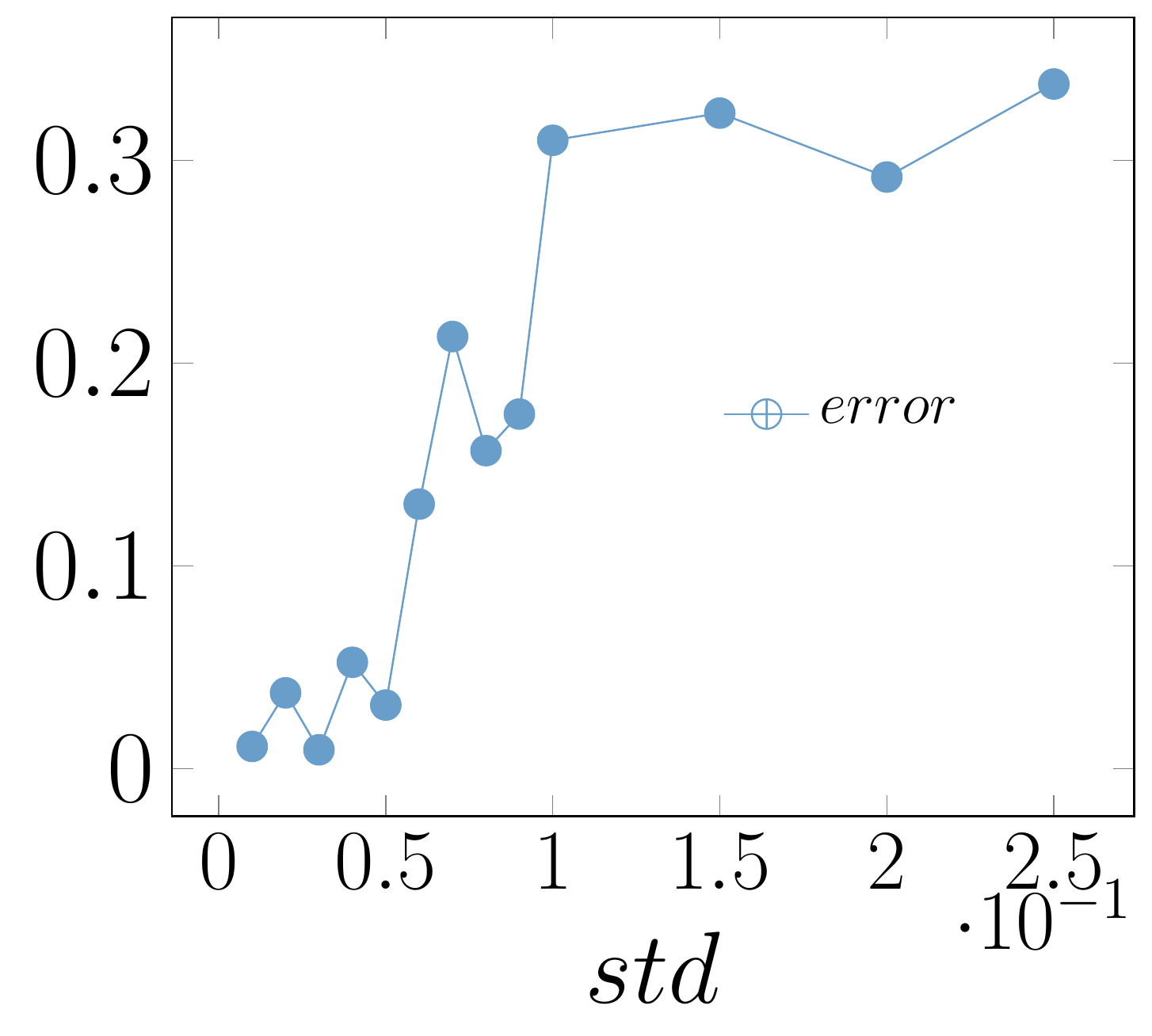}
}
\caption{
(a) The entropy of system $B$ of the 6-qubit Ising model with $J = 1$ and $\beta = 1$ as functions of the evolution time, the time duration $\delta t = 1$. 
(b) The final entropy of $B$ as the function of $J$ for Ising model with fixed parameter $ \beta = 1$.  $S_i$ is the initial entropy of $\rho_B$,and $S_t$ is the theoretical lower bound of $S(\rho_B(T))$.
(c) The final entropy of $B$ as the function of temperature $\beta$ for Ising model with fixed parameter $ J = 1 $. 
(d) The relative error of the final entropy with imperfect timekeeping evolution as functions of variance of the timekeeping tick $\sigma$ for a 6-qubit Ising model with $dt = 0.5$.}
\label{entropy_ising}
\end{figure}

In this section we show some numerical results about our entropy minimization protocol in the spin~(qubit) system.
In our previous paper~\cite{an2022learning}, the theoretical lower bound of one subsystem entropy in spin(qubit) system was given. For a bipartite state $\rho_{AB}^0$ with the eigen-decomposition is $\rho_{AB}^0 = \sum_{j=1}^{d_{AB}} p_j |\psi_j\rangle \langle \psi_j|$, assuming these eigenvalues are in decreasing order, by performing a unitary operation $U$ on $AB$, the minimal entropy of the subsystem $B$ is
\begin{align} \label{previous_result}
    \min_{U} S(\rho_B(U) ) = -\sum_j p_j \log_2 p_j, 
\end{align}
where $q_j = \sum_{a=1}^{d_A} p_{(j-1)d_A + a} $.

Here we show the numerical results of our entropy minimization protocol on the transverse field Ising model, whose Hamiltonian is 
\begin{align}
    H_0 &= -\sum_{i=1}^L  J \sigma ^x_i \sigma^x_{i+1} - \sigma ^z_i,
\end{align}
and the control Hamiltonian is 
\begin{align}
    H_c(t) &= \gamma (t) J \sigma^x_{l_A} \sigma^x_{l_A+1}.
\end{align}

In Fig.~\ref{entropy_ising}(a) we plot the entropy of $\rho_B$, $\rho_B$, and the mutual information between $A$ and $B$, as the function of $t$. Here we choose the $\mathcal{C}(\gamma) = \{1, 0.5, 0.3, 0.2, 0.1, 0\}$, and set $L=6$, $l_A = 1$. The mutual information between $A$ and $B$, $I_{AB}$ first fluctuates, and decreases to nearly $0$ at around $t=50$, then increases at around $t=100$, finally decreases to $0$ in the end. We note that the time period of the $I_{AB}$ decreases exactly corresponds to the decreasing of $S_{A}$, which indicates that $I_{AB}$ acts as a kind of ``potential" to decrease $S_{B}$. In the landscape of $S_{B}$, $I_{AB}$ decrease to $0$ corresponding to getting a local minimum of $S_{B}$, and the increasing of $I_{AB}$ corresponding to climbing out of the local minimum. The entropy of $A$ reaches to the maximum value $1$ for $A$ is one qubit. 

In Fig.~\ref{entropy_ising}(b) and (c), we plot the $S(\rho_B(T))$ as the function of $\beta$ and $J$. When the parameter $J$ is relatively big, i.e., the two-body interaction in Hamiltonian is strong, the evolution time becomes longer. Strong interaction makes the initial ground state tends to be a GHZ state $(|+\rangle^{\otimes L} +|-\rangle^{\otimes L})/\sqrt{2}$, thus the initial state is strong entangled and the initial $I_{AB}$ is big, which leads to a longer evolution time. When the two-body interaction in Hamiltonian is very weak, the ground state tends to be a product state $|0\rangle^{\otimes L}$, the entropy minimization is nearly failed. Weak interaction leads to a small $I_{AB}$ of initial state and a weak ability to increase the $I_{AB}$ during the evolution, thus leads to no such kind of ``potential" to decrease $S_{B}$. 

In Fig.~\ref{entropy_ising}(d) we show the relative error of the final entropy with imperfect timekeeping evolution as functions of variance of the timekeeping tick $\sigma$ with $dt$ is $0.5$. Our protocol retains valid when the variance of the timekeeping tick distribution $\sigma$ is lower than $10\%$ of $dt$.

\begin{table*}[htbp]
  \centering
  \begin{tblr}{|c|ccccccc|}
    \hline
    N & $4$ &  $5$ &  $6$ &  $7$ &  $8$ &  $10$&\\
    \hline
    error &0.013&0.092&0.019&0.065&0.047&0.044&\\
    \hline
    total time &30&500&300&420&500&220&\\
    \hline
  \end{tblr}
  \caption{The difference between the final entropy and the theoretical lower bound, and total evolution time for different size of Ising models. The total time and difference are model-dependent and parameter-dependent. The total time is relatively scaling as $O(1)$ and is not scaling exponentially with $N$.}\label{tab:ising}
\end{table*}

\end{document}